\documentclass{IEEEtran}
\usepackage{subfigure}
\usepackage[cmex10]{amsmath}
\usepackage{amssymb}
\usepackage{epsfig}
\usepackage{cite}
\usepackage{graphicx}
\usepackage{color}
\usepackage{bm}
\usepackage{booktabs}
\usepackage{gensymb}
\usepackage{mathrsfs}
\usepackage{xfrac}
\usepackage{algorithm}
\usepackage{algorithmic}
\usepackage{multirow}
\usepackage{float}
\newfloat{figtab}{htb}{fgtb}
\makeatletter
 \newcommand\figcaption{\def\@captype{figure}\caption}
  \newcommand\tabcaption{\def\@captype{table}\caption}

\newlength{\figwidth}
\newcommand{\tabincell}[2]{\begin{tabular}{@{}#1@{}}#2\end{tabular}}



\begin{document}
\title{Near Field Communications for \\ DMA-NOMA Networks}
\author{Zheng Zhang,~\IEEEmembership{Graduate Student Member,~IEEE},  Yuanwei Liu,~\IEEEmembership{Fellow,~IEEE}, \\ Zhaolin Wang,~\IEEEmembership{Graduate Student Member,~IEEE}, Jian Chen,~\IEEEmembership{Member,~IEEE}, and Dong In Kim,~\IEEEmembership{Fellow,~IEEE}
\thanks{Zheng Zhang and Jian Chen are with the School of Telecommunications Engineering, Xidian University, Xi'an 710071, China (e-mail: zzhang\_688@stu.xidian.edu.cn; jianchen@mail.xidian.edu.cn).}
\thanks{Yuanwei Liu and Zhaolin Wang are with the School of Electronic Engineering and Computer Science, Queen Mary University of London, London E1 4NS, U.K. (e-mail: yuanwei.liu@qmul.ac.uk; zhaolin.wang@qmul.ac.uk;).}
\thanks{Dong In Kim is with the Department of Electrical and Computer Engineering, Sungkyunkwan University, Suwon 16419, South Korea (e-mail: dikim@skku.ac.kr).}}

\maketitle

\begin{abstract}
    A novel near-field transmission framework is proposed for dynamic metasurface antenna (DMA)-enabled non-orthogonal multiple access (NOMA) networks. The base station (BS) exploits the hybrid beamforming to communicate with multiple near users (NUs) and far users (FUs) using the NOMA principle. Based on this framework, two novel beamforming schemes are proposed. 1) For the case of the grouped users distributed in the same direction, a beam-steering scheme is developed. The metric of beam pattern error (BPE) is introduced for the characterization of the gap between the hybrid beamformers and the desired ideal beamformers, where a two-layer algorithm is proposed to minimize BPE by optimizing hybrid beamformers. Then, the optimal power allocation strategy is obtained to maximize the sum achievable rate of the network. 2) For the case of users randomly distributed, a beam-splitting scheme is proposed, where two sub-beamformers are extracted from the single beamformer to serve different users in the same group. An alternating optimization (AO) algorithm is proposed for hybrid beamformer optimization, and the optimal power allocation is also derived. Numerical results validate that: 1) the proposed beamforming schemes exhibit superior performance compared with the existing imperfect-resolution-based beamforming scheme; 2) the communication rate of the proposed transmission framework is sensitive to the imperfect distance knowledge of NUs but not to that of FUs.

\end{abstract}
\IEEEpeerreviewmaketitle\vspace{-5mm}

\section{Introduction}
Fuelled by the explosive growth of ubiquitous wireless communications and various intelligent applications, such as extended reality (XR), auto-driving, and Internet-of-Everything (IoE), the development of the next generation multiple access (NGMA) techniques for future wireless networks becomes imminent \cite{Y.Liu_NGMA,Z.Ding_NGMA,J.Che_NGMA}. To enable flexible and reliable access to the network for a massive amount of users, each generation of multiple-access technique is committed to exploiting the new multiplexing domain schemes, such as the frequency-domain-based first-generation (1G) frequency division multiple access (FDMA), time-domain-based second-generation (2G) time division multiple access (TDMA) \cite{TDMA}, code-domain-based third-generation (3G) code division multiple access (CDMA) \cite{CDMA}, and orthogonal-subcarrier-based forth-generation (4G) orthogonal frequency division multiple access (OFDMA) \cite{OFDMA}. However, due to the limited spectrum resources, the aforementioned orthogonal multiple access (OMA) schemes are struggling to accommodate massive wireless connectivity. To deal with this challenge, the power-domain multiplexing-based non-orthogonal multiple access (NOMA) technique has drawn extensive attention in recent years \cite{Islam_NOMA}. By leveraging the superposition coding (SC) and successive interference cancellation (SIC) at the transmitters and receivers respectively, NOMA allows to serve multiple users within the same spectrum resource. The superiority of NOMA technology lies not only in its higher spectrum and energy efficiency gains compared to OMA schemes \cite{Y.Liu_NOMA_proceedings}, but also in its compatibility, which can be flexibly integrated into existing OMA communication systems. Especially in recent research, it has been claimed that NOMA can be utilized as an add-on to the conventional space division multiple access (SDMA) further to enhance the connectivity and spectral efficiency of multi-antenna networks \cite{NF_NOMA2,NF_NOMA3}.

Courtesy of the rapid development of metamaterials, a new antenna paradigm, namely dynamic metasurface antenna (DMA), has been proposed. Depending on the Lorentz resonance response characteristics of each element, it can be classified into two categories, i.e., amplitude-control DMA versus Lorentzian-constrained phase-shift control DMA \cite{DMA}. For ease of hardware implementation, the amplitude-control DMA (also referred to as reconfigurable holographic surface \cite{DMA_AP}) is considered in this paper. To elaborate, the DMA utilizes reference electromagnetic (EM) waves generated by feeds, which propagate along a metasurface inscribed with the beam pattern and radiate from a radiating element into free space \cite{DMA_WCM}. By recording the interference between the reference wave and the desired wave, the amplitude of the reference wave can be precisely controlled to generate the reconfigurable beam pattern, thus realizing the spatial beamformer \cite{DMA_AP2}. Compared to the conventional phased array antennas, DMA does not rely on the active amplifier and the phase-shift circuits, thus having lower energy consumption and hardware implementation cost. As an emerging antenna solution for wireless communications, a few works have been devoted to the beam pattern design in DMA-aided wireless networks \cite{R.Deng_DMA1,R.Deng_DMA2,X.Zhang_DMA,J.Hu_DMA}. Specifically, the authors of \cite{R.Deng_DMA1} proposed a DMA-enabled downlink multi-user transmission framework, where a hybrid beamforming design was devised to realize accurate multi-beam-steering. Followed by this, the authors of \cite{R.Deng_DMA1} proposed a new multiple access scheme, namely holographic-pattern division multiple access (HDMA), which was demonstrated to exhibit a higher network capacity than the conventional SDMA. In  \cite{X.Zhang_DMA}, a DMA-empowered holographic radar was developed for target sensing, which consumed less power than the phased array-based radar under the same sensing accuracy requirement. Moreover, the authors of \cite{J.Hu_DMA} proposed to exploit the multi-band DMA for user positioning, where the federated learning (FL) framework was adopted to improve sensing adaptability while guaranteeing privacy.

Inspired by the advantages of DMA, it is natural to focus on the investigation of DMA-enabled multi-user networks from the multiple access perspective, where the power-domain-based NOMA is considered to be integrated into DMA transmission to further improve spectral efficiency. Nevertheless, the low hardware cost benefit of DMA also results in the fact that antenna arrays tend to be extremely large in DMA networks, e.g., hundreds or even thousands of antennas could be deployed at the base station (BS). Such an enlargement of the array antenna scale leads to a fundamental propagation characteristic shift of the EM wave, which might change the communication range from the far-field EM radiated region to the near-field EM radiated region. Specifically, in far-field regions, the EM propagation can be approximated as the planar wave, where the planar-wave-based linear phase response should be adopted. In near-field regions, the EM propagation follows the more complex spherical wave, where the spherical-wave-based non-linear phase response with respect to both angle and distance knowledge is required to characterize the signal propagation \cite{X.Mu_NF}. Compared with the far-field channels, the near-field channel model introduces an extra distance dimension knowledge to favor wireless transmission design \cite{H.Zhang_NF,ZZ_NF_PLS,Z.Wang_NF}. To elaborate, the authors of \cite{H.Zhang_NF} proposed to exploit the distance information to achieve the signal power focusing on the desired location of free space, (referred as to \textit{beamfocusing}), which reduced the leakage of beam energy at uninterested locations and improved spectral efficiency. The authors of \cite{ZZ_NF_PLS} revealed the distance-domain secrecy gain brought by the near-field channels. In the work \cite{Z.Wang_NF}, the coupled angle and distance information implicit in the spherical-wave channels was used to achieve simultaneous angle and distance sensing. More recently, several preliminary studies have been devoted to exploring the possibilities of NOMA in near-field transmission by using it as an add-on to SDMA \cite{NF_NOMA1,NF_NOMA2,Zuo_NF_NOMA,NF_NOMA3}. In particular, in the work \cite{NF_NOMA3}, the authors unveiled that the near-field beamfocusing resolution is always imperfect even though the number of antennas at the BS end tends to infinity, which indicates that any single near-field beamfocusing beamformer leaks power to other users as well, which provides theoretical backing for the application of NOMA in the near field.

\subsection{Motivations and Contributions}
Although there have been some preliminary studies oriented towards the design of SDMA/HDMA-based DMA multi-user transmission \cite{R.Deng_DMA1,R.Deng_DMA2,R.Deng_DMA3}, research on exploiting NOMA in DMA networks is in its infancy. Actually, since active modules at the BS, e.g., radio frequency (RF) chains and digital baseband processing modules, still face expensive hardware costs (especially in the millimeter wave or terahertz bands), the active module cannot achieve a one-to-one antenna match (i.e., the fully-digital architecture) in practice, which limits the capacity of DMA to serve multiple users. Generally, the number of users that the BS can support in the spatial domain is constrained by the number of RF chains. Fortunately, integrating NOMA into the SDMA technique provides a new solution for serving more users with limited RF chains. On the other hand, the spherical-wave-based near-field channels caused by the large-scale array of the DMA also pose new design challenges for applying NOMA in DMA networks. To elaborate, distance-domain knowledge contained in near-field channels introduces a new beam characteristic, i.e., beamfocusing \cite{H.Zhang_NF}, which implies that the near-field beam width is narrower than that of the far-field beam. In particular, even the users located in the same direction cannot be covered by a single beam as in the far-field case. Although it has been rigorously proved that beam focussing is unlikely to be of perfect resolution \cite{NF_NOMA1,NF_NOMA2,Zuo_NF_NOMA,NF_NOMA3}, the fact that far-field users receive much less power than near-field users makes it difficult to ensure fairness in NOMA transmissions.

Against the above discussion, this paper focuses on an overloaded communication scenario (with much more communication users than RF chains,), and proposes a DMA-enabled near-field NOMA framework. Our goal is to maximize network capacity while guaranteeing fairness between near and far users through dedicated near-field beamformer design. The main contributions of this work are summarized below.
\begin{itemize}
  \item We propose a DMA-enabled near-field NOMA communication framework, where a BS exploits the hybrid DMA architecture to send signals to multiple near users (NUs) and far users (FUs) in a NOMA principle. Considering the limitation of the number of RF chains, each NU is associated with an FU to form a NOMA group, where a shared hybrid beamformer is designed for each group. Based on the designed beamformers, a power allocation optimization problem is formulated to maximize the sum achievable rate under the QoS requirement and the SIC decoding constraint.
  \item We first consider a special user location topology in which two users belonging to the same group are located in the same direction but at different distances. A beam-steering beamforming scheme is proposed, which aims to generate large beam-depth beamformers to radiate the same signal power at the locations of the NUs and FUs for fairness. Specifically, we introduce the metric of the beam pattern error (BPE) to evaluate the gap between the practical beamformers and the desired ideal beamformers. A BPE minimization problem is formulated.  Then, we propose a two-layer algorithm to iteratively optimize the hybrid beamformers. On this basis, the optimal power allocation strategy is derived to enhance the spectral efficiency of the network.
  \item We further consider a general case with randomly distributed users, where a beam-splitting beamforming scheme is proposed. To elaborate, we decompose the original beamformer into two sub-beamformers, which are used to serve the different users within the same group, respectively. To guarantee user fairness, a minimum channel gain maximization is maximized under the amplitude constraint of the DMA elements and inter-group interference limitation. To this end, an alternating optimization (AO) algorithm is proposed to optimize the hybrid beamformers. Then, the optimal power allocation is obtained for sum-rate maximization.
  \item Simulation results verify the convergence of the proposed algorithms. It is also found that: 1) the proposed beam-steering and -splitting schemes outperform the existing imperfect-resolution-based single beam scheme in near-field NOMA transmission; 2) the communication performance of the proposed DMA-enabled near-field NOMA transmission framework is sensitive to the imperfect distance knowledge of NU, but virtually unaffected by that of the FU.
\end{itemize}

\subsection{Organization and Notations}\vspace{-1mm}
The remainder of this paper is as follows. Section \ref{System Model} introduces the network setup and signal model. Section \ref{beam-steering} proposes a two-layer algorithm for the beam-steering beamformer design. Section \ref{beam-splitting} conceives an AO algorithm for the beam-splitting beamformer design. We provide the numerical results in Section \ref{Numerical Results}. The conclusion is drawn in Section \ref{Conclusion}.

\textit{Notations:} The scalar, vector, and matrix are represented by the lower-case letter, boldface lower-case letter, and boldface capital, respectively. The transpose and Hermitian conjugate operations of matrix $\mathbf{X}$ are denoted by $\mathbf{X}^{T}$ and $\mathbf{X}^{H}$. The $i$-th row and $j$-th column element of the matrix $\mathbf{X}$ is denoted by $\mathbf{X}^{[i,j]}$. The Euclidean norm of the vector $\mathbf{x}$ is denoted by $\|\mathbf{x}\|$. The circularly symmetric complex Gaussian (CSCG) distributed random variable with zero mean and covariance matrix $a$ is denoted by $x\sim \mathcal{CN}(0,a)$. $\text{Tr}(\mathbf{X})$, $\text{rank}(\mathbf{X})$, and $(\mathbf{X})^{-1}$ denote the trace, rank and inverse-matrix operations. $\mathbf{X}\succeq \mathbf{0}$ represents that $\mathbf{X}$ is a semi-definite matrix. $\Re({\cdot})$ denotes the real component of the corresponding complex value. $\jmath$ denotes the unit imaginary number.

\section{System Model}\label{System Model}

\subsection{Network Description}
\begin{figure}[t]
  \centering
  \includegraphics[scale = 0.35]{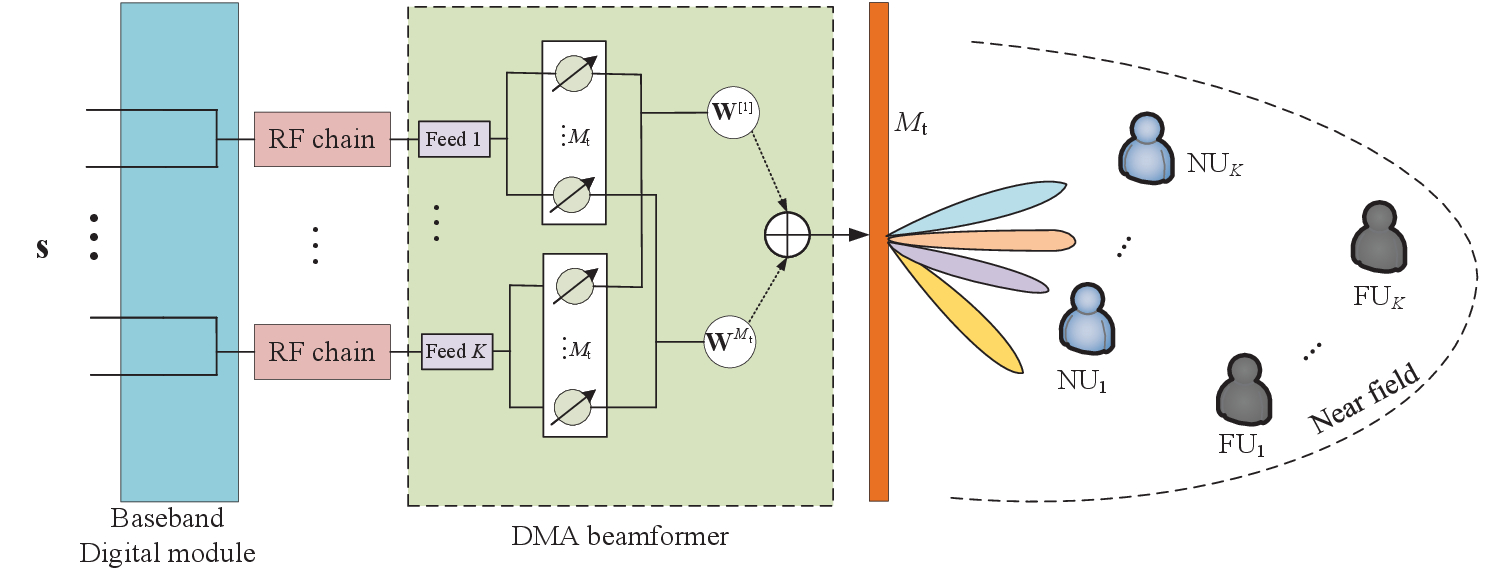}
  \caption{Hybrid DMA-enabled near-field NOMA communications.}
  \label{Fig.1}
\end{figure}
In this paper, we consider a downlink multi-user network, where a BS communicates with $2K$ users by utilizing the radiation pattern of the DMA. The DMA is equipped with $M_{\text{t}}=M_{\text{t},v}M_{\text{t},h}$ elements, where $M_{\text{t},v}$ and $M_{\text{t},h}$ denote the number of elements located in the vertical and horizontal directions of the DMA, respectively. As shown in Fig. \ref{Fig.1}, two categories of users are considered in the network, where $K$ near users (denoted by $\{\text{NU}_{1},\cdots,\text{NU}_{K}\}$) are located in the vicinity of the BS while the far $K$ users (denoted by $\{\text{FU}_{1},\cdots,\text{FU}_{K}\}$) lie relatively far away from the BS. All the users are assumed to be single-antenna nodes, each of which only requires a single data stream from the BS. Under the extremely large-scale DMA setup, we assume all the users are located in the Fresnel (near-field) region of the BS, which implies that the distances between the users and the BS are shorter than the Rayleigh distance $\frac{2D^{2}}{\lambda}$, with $D$ representing the aperture of the BS.

This paper focuses on a connectivity-overloaded network. To elaborate, $M_{\text{RF}}$ ($K\leq M_{\text{RF}}< 2K$) RF chains are integrated into the BS to connect with $M_{\text{RF}}$ independent feeds of the DMA, each of which is capable of generating EM waves as the incident signals. For simplification, we assume that $M_{\text{RF}} = K$ in the following, i.e., the BS allows the generation of up to $K$ independent digital beams to serve different users. To further increase the network connectivity with the limited hardware overhead of the RF chains, the NOMA technique is exploited to serve more users with the same spatial DoF. In particular, $2K$ users are clustered as $K$ NOMA groups, each of which consists of one NU and one FU. By employing the superposition coding (SC) at the BS and the successive interference cancellation (SIC) at users, the NOMA protocol is adopted within each group for multiple data transmission.

\subsection{DMA Architecture}
Unlike the conventional phase-array-based antennas, DMA is a category of planar antenna, which avoids the deployment of the power amplifiers and phase shifters, thus resulting in low energy costs. From the hardware architecture perspective, DMA is generally composed of three parts, i.e., feeds, waveguides, and metamaterial radiation elements. Specifically, the feeds are deployed at the bottom layer of the DMA, which are responsible for receiving the incident signals up-converted by the RF chains and generating the reference EM wave. To guide the wave propagation, the waveguide is distributed along the DMA surface, which serves as the propagation medium of the reference EM waves and radiates reference EM waves into free space. On the top layer of the DMA, the metamaterial radiation elements are mounted, which can intelligently control the radiation pattern of the reference EM waves by altering its EM response at each element.

The construct of the beamformer at the DMA relies on the physical interference principle of the EM wave. To elaborate, by recording the interference pattern between the desired wave and the reference EM wave, the DMA can generate the radiation pattern that orientates towards the direction of interest. Let $\mathbf{x}_{v,h}$ denotes the location vector of the $v$-th row and the $h$-th column element, the reference wave activated by the feed $k$ with respect to the $v$-th row and the $h$-th column element is given by
\begin{align}\label{DMA_ref_wave}
\Gamma_{\text{r}}(\mathbf{x}_{v,h}^{k},\mathbf{r}_{\text{s}}) = e^{-\jmath \mathbf{r}_{\text{s}} \mathbf{x}_{v,h}^{k}},
\end{align}
where $\mathbf{r}_{\text{s}}$ denotes the propagation vector of the reference EM wave and $\mathbf{x}_{v,h}^{k}$ denotes the location information vector between the feed $k$ and the $v$-th row and the $h$-th column element. Similarly, the objective wave with respect to the $v$-th row and the $h$-th column element is given by
\begin{align}\label{DMA_obj_wave}
\Gamma_{\text{o}}(\mathbf{x}_{v,h},\mathbf{r}_{\text{f}}) = e^{-\jmath \mathbf{r}_{\text{f}} \mathbf{x}_{v,h}}.
\end{align}
Here, $\mathbf{r}_{\text{f}}$ denotes the propagation vector from the origin of the coordinate system to the objective location $(\phi,\varphi,r)$, in which $\phi$, $\varphi$, and $r$ denote the corresponding azimuth angle, elevation angle, and the distance information of the objective position. Therefore, the interference pattern between $\Gamma_{\text{r}}(\mathbf{x}_{v,h}^{k},\mathbf{r}_{\text{s}})$ and $\Gamma_{\text{o}}(\mathbf{x}_{v,h},\mathbf{r}_{\text{f}})$ can be expressed as
\begin{align}\label{DMA_int_wave}
\Gamma_{\text{i}}(\mathbf{x}_{v,h}^{k},\mathbf{r}_{\text{f}}) = \Gamma_{\text{r}}(\mathbf{x}_{v,h}^{k},\mathbf{r}_{\text{s}})\Gamma_{\text{o}}^{*}(\mathbf{x}_{v,h},\mathbf{r}_{\text{f}}),
\end{align}
where it is readily verified that the radiation pattern excited by the reference EM wave orientates the direction of the objective position, i.e., $\Gamma_{\text{i}}(\mathbf{x}_{v,h},\mathbf{r}_{\text{f}})\Gamma_{\text{r}}(\mathbf{x}_{v,h}^{k},\mathbf{r}_{\text{s}})
\propto \Gamma_{\text{o}}(\mathbf{x}_{v,h},\mathbf{r}_{\text{f}}) C_{v,h}^{k}$ ($C_{v,h}^{k}$ is a constant equals $C_{v,h}^{k}=|\Gamma_{\text{r}}(\mathbf{x}_{v,h}^{k},\mathbf{r}_{\text{s}})|^{2}$). To adjust the interference radiation pattern, an amplitude-variation-based controlling approach is adopted at the DMA, where the normalized radiation amplitude for the feed $k$ at the $v$-th row and the $h$-th column element is given by
\begin{align}\label{DMA_amp_contr}
\mathbf{W}^{[(v-1)h+h,k]} = \frac{\Re{(\Gamma_{\text{i}}(\mathbf{x}_{v,h}^{k},\mathbf{r}_{\text{f}})})+1}{2}e^{-\jmath \mathbf{r}_{\text{s}} \mathbf{x}_{v,h}^{k}}.
\end{align}
Based on the above, we consider a hybrid DMA-assisted multiuser transmission architecture, where the baseband digital module emits signals intended for multiple users through the baseband digital beamformer. Then, $K$ data streams are sent to the feeds of the DMA via $K$ RF chains in parallel, which are further manipulated by the beamformer $\mathbf{W}\in\mathbb{C}^{M_{\text{t}}\times K}$ for signal broadcasting.

\subsection{Near-Field Channel Model}
\begin{figure}[t]
  \centering
  \includegraphics[scale = 0.6]{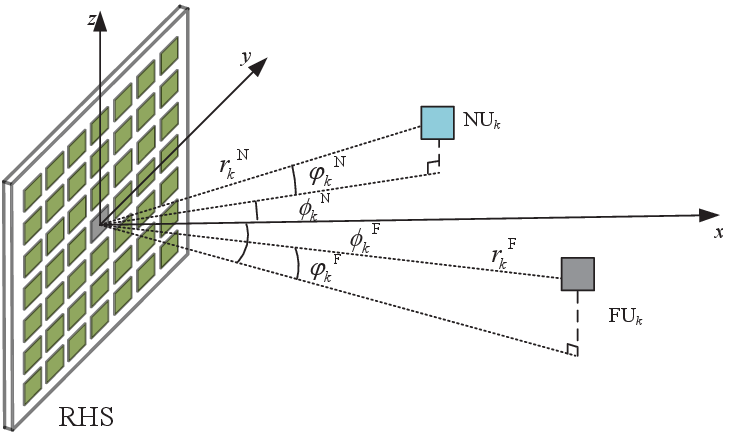}
  \caption{Near-field channel model.}
  \label{Fig.2}
\end{figure}

As all the users are located in the near-field region of the network, the accurate spherical-wave channel model is required. Note that as the high-frequency (e.g., millimeter wave or THz) channel is generally predominated by the LoS component, we only consider the LoS channel in this paper. As depicted in Fig. \ref{Fig.2}, consider a three-dimensional (3D) coordinate system, with the DMA being located in the y-o-z plane. With the assumption that the central element of the DMA is located at the origin of the coordinate system, the $v$-th row and the $h$-th column element is located in the coordinate of $\mathbf{x}_{v,h}=(0,\tilde{v}d,\tilde{h}d)$, where $\tilde{v}=v-\frac{M_{\text{t},v}+1}{2}$, $\tilde{h}=h-\frac{M_{\text{t},h}+1}{2}$, $d=\frac{\lambda}{2}$ denotes the inter-distance between the adjacent elements, and $\lambda$ denotes the wavelength of carrier wave. Thus, the Euclidean distance between the the $v$-th row and the $h$-th column element of the DMA and the $\text{NU}_{i}$/$\text{FU}_{i}$ is given by
\begin{align}\nonumber
&\|\mathbf{x}_{v,h} - \mathbf{s}_{i}^{\varsigma}\|=  \big[(r_{i}^{\varsigma})^{2}+\tilde{v}^{2}d^{2}+\tilde{h}^{2}d^{2}-2r_{i}^{\varsigma}\\ \label{NF_distance}
&\qquad
\tilde{v}d\sin\phi_{i}^{\varsigma}\sin\varphi_{i}^{\varsigma}-2r_{i}^{\varsigma}
\tilde{h}d\cos\varphi_{i}^{\varsigma}\big]^{\frac{1}{2}},\  \varsigma\in\{\text{N},\text{F}\},
\end{align}
where $\mathbf{s}_{i}^{\varsigma}=(r_{i}^{\varsigma}\cos\phi_{i}^{\varsigma}\sin\varphi_{i}^{\varsigma},
r_{i}^{\varsigma}\sin\phi_{i}^{\varsigma}\sin\varphi_{i}^{\varsigma},
r_{i}^{\varsigma}\cos\varphi_{i}^{\varsigma})$ denotes the coordinate of the user. Here, the distance from the origin of the coordinate system to the $\text{NU}_{i}$/$\text{FU}_{i}$ is denoted by $r_{i}^{\varsigma}$, the azimuth and elevation angles are represented as $\phi_{i}^{\varsigma}$ and $\varphi_{i}^{\varsigma}$, respectively. Accordingly, the array response vector from the BS to the $\text{NU}_{i}$/$\text{FU}_{i}$ can be expressed as
\begin{align} \label{LoS_arr_res}
\mathbf{a}^{\varsigma}_{i} = \big[e^{-\jmath\frac{2\pi}{\lambda}(\|\mathbf{x}_{1,1} - \mathbf{s}_{i}^{\varsigma}\|)},\cdots,e^{-\jmath\frac{2\pi}{\lambda}(\|\mathbf{x}_{M_{\text{t},\text{v}},M_{\text{t},\text{h}}} - \mathbf{s}_{i}^{\varsigma}\|)}\big]^{T}.
\end{align}
Thus, the channel between the BS and the $\text{NU}_{i}$/$\text{FU}_{i}$ can be expressed as $\mathbf{h}^{\varsigma}_{\text{LoS},i}=\beta_{i}e^{-\jmath\frac{2\pi}{\lambda}r_{i}^{\varsigma}}\mathbf{a}^{\varsigma}_{i}$, where $\beta_{i}^{\varsigma}$ denotes the complex gain defined in \cite{Y.Liu_NFC_tutorial}. Furthermore, we assume that the perfect CSI of near-field users is known at the BS.

\subsection{Signal Model for NOMA Transmission}
To enhance the connectivity of the DMA-enabled network, a NOMA-empowered transmission scheme is proposed. To elaborate, we consider cluster $2K$ users as $K$ NOMA pairs, where each NOMA pair is associated with one RF chain and consists of a NU and a FU. Within a NOMA pair, the joint design of the hybrid beamformers and the power allocation are performed, where the hybrid beamformers are required to be deliberately generated to align the superimposed NOMA signals to the locations of users. For simplification, we assume that the $i$-th NOMA group is composed of $\text{NU}_{i}$ and $\text{FU}_{i}$ ($1\leq i\leq K$). Accordingly, the emitted signals from the BS can be expressed by
\begin{align} \label{tra_sig}
\mathbf{x} = \sum_{i=1}^{K}\mathbf{W}\mathbf{v}_{i}\left(\sqrt{P_{1,i}}s_{i}^{\text{N}}+\sqrt{P_{2,i}}s_{i}^{\text{F}}\right),
\end{align}
where $\mathbf{v}_{i}$ is the baseband digital beamforming vector allocated to $i$-th NOMA group, $s_{i}^{\text{N}}$ and $s_{i}^{\text{F}}$ denote the signals intended for the $\text{NU}_{i}$ and $\text{FU}_{i}$, respectively, with satisfying $\mathbb{E}\{|s_{i}^{\text{N}}|^{2}\}=\mathbb{E}\{|s_{i}^{\text{F}}|^{2}\}=1$. Note that $P_{1,i}$ and $P_{2,i}$ are the transmit power allocated to the $\text{NU}_{i}$ and $\text{FU}_{i}$. Thus, the received signals at the $\text{NU}_{i}$ and $\text{FU}_{i}$ are given by
\begin{align} \nonumber
\mathbf{y}_{i}^{\text{N}} = &\underbrace{\mathbf{h}^{\text{N}}_{i}\mathbf{W}\mathbf{v}_{i}\sqrt{P_{1,i}}s_{i}^{\text{N}}}_{\text{desired signal}}+\underbrace{\mathbf{h}^{\text{N}}_{i}\mathbf{W}\mathbf{v}_{i}\sqrt{P_{2,i}}s_{i}^{\text{F}}}_{\text{intra-group interference}}+\\ \label{sig_NU}
& \underbrace{\mathbf{h}^{\text{N}}_{i}\sum_{t=1,t\neq i}^{K}\mathbf{W}\mathbf{v}_{t}\left(\sqrt{P_{1,t}}s_{t}^{\text{N}}+\sqrt{P_{2,t}}s_{t}^{\text{F}}\right)}_{\text{inter-group interference}}+n_{i}^{\text{N}},
\end{align}
\begin{align} \nonumber
\mathbf{y}_{i}^{\text{F}} = &\underbrace{\mathbf{h}^{\text{F}}_{i} \mathbf{W}\mathbf{v}_{i}\sqrt{P_{2,i}}s_{i}^{\text{F}}}_{\text{desired signal}}+\underbrace{\mathbf{h}^{\text{F}}_{i} \mathbf{W}\mathbf{v}_{i}\sqrt{P_{1,i}}s_{i}^{\text{N}}}_{\text{intra-group interference}}+\\ \label{sig_FU}
& \underbrace{\mathbf{h}^{\text{F}}_{i}\sum_{t=1,t\neq i}^{K} \mathbf{W}\mathbf{v}_{t}(\sqrt{P_{1,t}}s_{t}^{\text{N}}+\sqrt{P_{2,t}}s_{t}^{\text{F}})}_{\text{inter-group interference}}+n_{i}^{\text{F}}.
\end{align}
Note that $n_{i}^{\text{N}},n_{i}^{\text{F}}\sim \mathcal{CN}(0,\sigma^{2})$ denote the additive white Gaussian noise (AWGN) at the $\text{NU}_{i}$ and $\text{FU}_{i}$, respectively. On receiving the superimposed signals, the SIC technique is employed at each NOMA group. To elaborate, the user with a strong channel condition first decodes the signal of the weak-channel user, and removes it from the received signal observation. Then, the strong-channel user decodes its own signal without suffering intra-group interference. For the user with a weak channel condition, it directly decodes its own signal by treating the strong-channel user signal as the intra-group interference. For our considered network, the NU is naturally treated as the strong-channel user and the FU is the weak-channel user. Therefore, the received signal-to-interference-plus-noise ratio (SINR) at $\text{NU}_{i}$ and $\text{FU}_{i}$ is given by
\begin{align} \label{SINR_NU}
\gamma_{\text{N}_{i}\rightarrow \text{N}_{i}} = \frac{P_{1,i}|(\mathbf{h}^{\text{N}}_{i})^{H}\mathbf{W}\mathbf{v}_{i}|^{2}}
{I_{i,\text{inter}}^{\text{N}}+\sigma^{2}},
\end{align}
\begin{align} \label{SINR_FU}
\gamma_{\text{F}_{i}\rightarrow \text{F}_{i}} = \frac{P_{2,i}|(\mathbf{h}_{i}^{\text{F}})^{H}\mathbf{W}\mathbf{v}_{i}|^{2}}
{P_{1,i}|(\mathbf{h}_{i}^{\text{F}})^{H}\mathbf{W}\mathbf{v}_{i}|^{2}+
I_{i,\text{inter}}^{\text{F}}+\sigma^{2}},
\end{align}
where $I_{i,\text{inter}}^{\text{N}}=\sum_{t=1,t\neq i}^{K}(P_{1,t}|(\mathbf{h}^{\text{N}}_{i})^{H}\mathbf{W}\mathbf{v}_{t}|^{2}+
P_{2,t}|(\mathbf{h}^{\text{N}}_{i})^{H}\mathbf{W}\\ \mathbf{v}_{t}|^{2})$ and $I_{i,\text{inter}}^{\text{F}}=\sum_{t=1,t\neq i}^{K}(P_{1,t}|(\mathbf{h}^{\text{F}}_{i})^{H}\mathbf{W}\mathbf{v}_{t}|^{2}+
P_{2,t} |(\mathbf{h}^{\text{F}}_{i})^{H}\\ \mathbf{W}\mathbf{v}_{t}|^{2})$. To ensure the successful SIC decoding procedure, it is also required that the achievable rate of $\text{NU}_{i}$ to decode $s_{i}^{\text{F}}$ should be no less than the achievable rate of $\text{FU}_{i}$ to decode its own signal, i.e., $\log_2(1+\gamma_{\text{N}_{i}\rightarrow \text{F}_{i}})\geq\log_2(1+\gamma_{\text{F}_{i}\rightarrow \text{F}_{i}})$, where $\gamma_{\text{N}_{i}\rightarrow \text{F}_{i}}$ is given by
\begin{align} \label{SINR_NUdFU}
\gamma_{\text{N}_{i}\rightarrow \text{F}_{i}} = \frac{P_{2,i}|(\mathbf{h}^{\text{N}}_{i})^{H}\mathbf{W}\mathbf{v}_{i}|^{2}}
{P_{1,i}|(\mathbf{h}_{i}^{\text{N}})^{H}\mathbf{W}\mathbf{v}_{i}|^{2}+
I_{i,\text{inter}}^{\text{N}}+\sigma^{2}}.
\end{align}

\section{beam-steering beamformer Design}\label{beam-steering}
\begin{figure}[t]
  \centering
  \includegraphics[scale = 0.6]{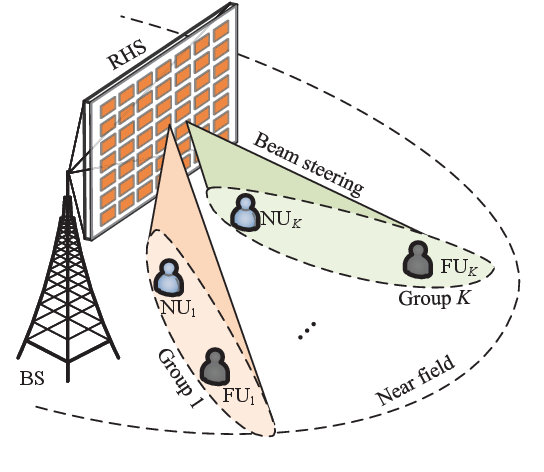}
  \caption{beam-steering scheme for near-field NOMA transmission.}
  \label{Fig.3}
\end{figure}

In this section, we consider a special user location topology, i.e., each FU is located in the same direction as its paired NU. For this scenario, we propose a beam-steering-based hybrid beamforming scheme, which simultaneously aligns with the NU and the FU in one NOMA group using the large beam-depth beamformer (see Fig. \ref{Fig.3}). Then, an optimal power allocation algorithm is proposed to maximize the network spectral efficiency.

\subsection{Large Beam-Depth Beamformer Design}
To restrict the beam-steering characteristics of the designed beamformer, we introduce a new performance metric, namely BPE, which aims to characterize the error degree of the designed beamformer compared with the required beam pattern \cite{B.Ning_wide_beam}. Based on the near-field array response vector $\mathbf{a}(\phi_{i},\varphi_{i},r_{i})$, the BPE for the beamformer $\{\mathbf{v}_{i},\mathbf{W}\}$ is defined by
\begin{align}\nonumber
\chi(\mathbf{v}_{i},\mathbf{W}) &\triangleq \int_{r\in[r^{\text{N}}_{i},r^{\text{F}}_{i}]}
\left|t-\left|\mathbf{a}(\phi_{i},\varphi_{i},r)^{H}\mathbf{W}\mathbf{v}_{i}\right|\right|^{2}dr+\\  \nonumber
&\int_{r\in[0,r^{\text{N}}_{i}]\cup[r^{\text{F}}_{i},\infty]}
\left|\mathbf{a}(\phi_{i},\varphi_{i},r)^{H}\mathbf{W}\mathbf{v}_{i}\right|^{2}dr+\\ \label{BPE_def}
&\iiint\limits_{\substack{{\phi\in[-\frac{\pi}{2},\phi_{i})\cup(\phi_{i},\frac{\pi}{2}],}\\ {\varphi\in[-\frac{\pi}{2},\varphi_{i})\cup(\varphi_{i},\frac{\pi}{2}]}\\
{r\in[0,\infty]}}}\!\!\!\!
\left|\mathbf{a}(\phi,\varphi,r)^{H}\mathbf{W}\mathbf{v}_{i}\right|^{2}d\phi d\varphi dr,
\end{align}
where $t$ denotes the desired strength of the ideal beam-steering vector. With the above definition, the beam-steering beamformer design problem can be formulated as
\begin{subequations}
\begin{align}
\label{P1a} (\text{P}1)\  \min\limits_{\mathbf{W},\mathbf{v}_{i}}\quad &   \sum_{i=1}^{K}\chi(\mathbf{v}_{i},\mathbf{W})\\
\label{P1b}\text{s.t.} \quad & \|\mathbf{v}_{i}\|^{2} = 1,\ 1\leq i\leq K,\\
\label{P1c}  &\mathbf{W}^{[m,n]}\in [0,1], \  1\leq m\leq M_{\text{t}},\ 1\leq n\leq K,\\
\label{P1d}  &|(\mathbf{a}_{j}^{\varsigma})^{H}\mathbf{W}\mathbf{v}_{i}|^{2} \leq \epsilon,\ i\neq j,
\end{align}
\end{subequations}
where \eqref{P1b} denotes the normalized power constraint of the transmit baseband digital beamformer, \eqref{P1c} represents the amplitude-control constraint at the DMA, and \eqref{P1d} is to limit inter-group interference below a negligible level $\epsilon\rightarrow 0$. Note that the problem (P1) is intractable to solve from the perspective of convex optimization due to the integral operations and infinite integral interval. To facilitate the solving of the problem, we consider transforming the problem (P1) into a discrete form. Specifically, we first limit the distance interval from $[0,\infty]$ to a finite interval $[0,r_{\text{max}}]$. Then, the continuous BPE function are quantized into $Q$ discrete values with $Q_{1}$ azimuth angle samples, $Q_{2}$ elevation angle samples, and $Q_{3}$ distance samples ($Q_{1}Q_{2}Q_{3}=Q$), where the $q$-th discrete location information is given by
\begin{align}\label{Quan_loc}
\begin{cases}\phi_{q_{1}} = -\frac{\pi}{2}+\frac{(q_{1}-1)\pi}{Q_{1}-1},\ q_{1}\in\{1,\cdots,Q_{1}\},\\
\varphi_{q_{2}} = -\frac{\pi}{2}+\frac{(q_{2}-1)\pi}{Q_{2}-1},\ q_{2}\in\{1,\cdots,Q_{2}\},\\
r_{q_{3}} = \frac{(q_{3}-1)r_{\text{max}}}{Q_{3}-1},\ q_{3}\in\{1,\cdots,Q_{3}\}, \\
\end{cases},
\end{align}
where there exists a unique mapping rule between the scalar $q$ and the vector $[q_{1},q_{2},q_{3}]$, i.e., $q=q_{3}+(q_{2}-1)Q_{3}+(q_{1}-1)Q_{3}Q_{2}$. Thus, we approximate the BPE function for $\{\mathbf{v}_{i},\mathbf{W}\}$ as
\begin{align}\nonumber
\chi(\mathbf{v}_{i},\mathbf{W}) \overset{(a)}{\approx} &\chi_{1}(\mathbf{v}_{i},\mathbf{W})\\ \nonumber
= &\sum_{r_{q_{3}}\in[r^{\text{N}}_{i},r^{\text{F}}_{i}]}
\left|t-\left|\mathbf{a}(\phi_{i},\varphi_{i},r_{q_{3}})^{H}\mathbf{W}\mathbf{v}_{i}\right|\right|^{2}\Delta r+\\  \nonumber
&\sum_{r_{q_{3}}\in[0,r^{\text{N}}_{i}]\cup[r^{\text{F}}_{i},r_{\text{max}}]}\!\!\!\!\!\!\!\!
\left|\mathbf{a}(\phi_{i},\varphi_{i},r_{q_{3}})^{H}\mathbf{W}\mathbf{v}_{i}\right|^{2}\Delta r+\\ \label{BPE_def_Quan}
&\!\!\! \sum_{\substack{{\phi_{q_{1}}\in[-\frac{\pi}{2},\frac{\pi}{2}],\phi_{q_{1}}\neq\phi_{i}}\\ {\varphi_{q_{2}}\in[-\frac{\pi}{2},\frac{\pi}{2}],\varphi_{q_{2}}\neq\varphi_{i}}\\
{r_{q_{3}}\in[0,r_{\text{max}}]}}}\!\!\!\!\!\!\!\!\!\!\!\!\!\!\!\!
\left|\mathbf{a}(\phi_{q_{1}},\varphi_{q_{2}},r_{q_{3}})^{H}\mathbf{W}\mathbf{v}_{i}\right|^{2}\Delta \phi \Delta \varphi \Delta r,
\end{align}
where the approximate equality sign $a$ strictly takes equality when $\Delta \phi\rightarrow 0$, $\Delta \varphi\rightarrow 0$, and $\Delta r\rightarrow 0$ satisfy. Accordingly, we can convert the objective function $\chi_{1}(\mathbf{v}_{i},\mathbf{W})$ as the form as follows.
\begin{align}\nonumber
\chi_{2}(\mathbf{v}_{i},\mathbf{W}) &= \frac{\chi_{1}(\mathbf{v}_{i},\mathbf{W})}{\Delta r}\\ \nonumber
&=\sum_{r_{q_{3}}\in[r^{\text{N}}_{i},r^{\text{F}}_{i}]}
\left|t-\left|\mathbf{a}(\phi_{i},\varphi_{i},r_{q_{3}})^{H}\mathbf{W}\mathbf{v}_{i}\right|\right|^{2}\\  \nonumber
&+\sum_{r_{q_{3}}\in[0,r^{\text{N}}_{i}]\cup[r^{\text{F}}_{i},r_{\text{max}}]}
\left|\mathbf{a}(\phi_{i},\varphi_{i},r_{q_{3}})^{H}\mathbf{W}\mathbf{v}_{i}\right|^{2}+\\ &\sum_{\substack{{\phi_{q_{1}}\in[-\frac{\pi}{2},\frac{\pi}{2}],\phi_{q_{1}}\neq\phi_{i}}\\ {\varphi_{q_{2}}\in[-\frac{\pi}{2},\frac{\pi}{2}],\varphi_{q_{2}}\neq\varphi_{i}}\\ \label{BPE_def_Quan_sim}
{r_{q_{3}}\in[0,r_{\text{max}}]}}}\!\!\!\!\!\!\!\!\!\!\!\!\!\!\!\!\!
\left|\mathbf{\bar{a}}(\phi_{q_{1}},\varphi_{q_{2}},r_{q_{3}})^{H}\mathbf{W}\mathbf{v}_{i}\right|^{2},
\end{align}
where $\mathbf{\bar{a}}(\phi_{q_{1}},\varphi_{q_{2}},r_{q_{3}}) = \sqrt{\Delta \phi \Delta \varphi\mathbf{a}(\phi_{q_{1}},\varphi_{q_{2}},r_{q_{3}})}$. To deal with the coupling between $\mathbf{v}_{i}$ and $\mathbf{W}$, we introduce an auxiliary variable $\mathbf{\bar{v}}_{i}$ with satisfying $\mathbf{\bar{v}}_{i}=\mathbf{W}\mathbf{v}_{i}$. We consider adopting the penalty-based optimization framework, where the equality constraint $\mathbf{\bar{v}}_{i}=\mathbf{v}_{i}\mathbf{W}$ is moved to the objective function as a penalty term. Hence, the problem (P2) is transformed into
\begin{subequations}
\begin{align}
\label{P2a} (\text{P}2)\  \min\limits_{\mathbf{W},\mathbf{v}_{i},\mathbf{\bar{v}}_{i}}\quad &  \sum_{i=1}^{K}\chi_{2}(\mathbf{\bar{v}}_{i})+\frac{1}{2\rho}\sum_{i=1}^{K}\|\mathbf{\bar{v}}_{i}-\mathbf{W}\mathbf{v}_{i}\|^{2}\\
\label{P2b}\text{s.t.} \quad& \eqref{P1b},\eqref{P1c},\eqref{P1d},
\end{align}
\end{subequations}
where the penalty term $\|\mathbf{\bar{v}}_{i}-\mathbf{W}\mathbf{v}_{i}\|^{2}\rightarrow 0$ when $\rho\rightarrow 0$. To handle the double modulus operation in $\chi_{2}(\mathbf{\bar{v}}_{i})$, we introduce \cite[Lemma 1]{B.Ning_wide_beam}, which equivalently converts $\chi_{2}(\mathbf{\bar{v}}_{i})$ as following tractable form
\begin{align}\nonumber
&\chi_{2}(\mathbf{\bar{v}}_{i}) =\chi_{3}(\mathbf{\bar{v}}_{i},\vartheta_{q_{3}}^{i})=\!\!\!\sum_{r_{q_{3}}\in[r^{\text{N}}_{i},r^{\text{F}}_{i}]}\!\!\!
\left|te^{\jmath \vartheta_{q_{3}}^{i}}-\mathbf{a}(\phi_{i},\varphi_{i},r_{q_{3}})^{H}\mathbf{\bar{v}}_{i}\right|^{2}\\  \label{BPE_Quan_fina}
&+\!\!\!\!\!\!\!\sum_{r_{q_{3}}\in[0,r^{\text{N}}_{i}]\cup\atop[r^{\text{F}}_{i},r_{\text{max}}]}\!\!\!\!\!\!\!\!
\left|\mathbf{a}(\phi_{i},\varphi_{i},r_{q_{3}})^{H}\mathbf{\bar{v}}_{i}\right|^{2}+\!\!\!\!\!\!\!\!\!\!\!\!\!\!
\sum_{\substack{{\phi_{q_{1}}\in[-\frac{\pi}{2},\frac{\pi}{2}],\phi_{q_{1}}\neq\phi_{i}}\\ {\varphi_{q_{2}}\in[-\frac{\pi}{2},\frac{\pi}{2}],\varphi_{q_{2}}\neq\varphi_{i}}\\
{r_{q_{3}}\in[0,r_{\text{max}}]}}}\!\!\!\!\!\!\!\!\!\!\!\!\!\!
\left|\mathbf{\bar{a}}(\phi_{q_{1}},\varphi_{q_{2}},r_{q_{3}})^{H}\mathbf{\bar{v}}_{i}\right|^{2}.
\end{align}
With these transformations, we reformulate the optimization problem (P2) as
\begin{subequations}
\begin{align}
\label{P3a} (\text{P}3)\  \min\limits_{\mathbf{W},\mathbf{v}_{i},\mathbf{\bar{v}}_{i},\vartheta_{q_{3}}^{i}}\quad &  \sum_{i=1}^{K}\chi_{3}(\mathbf{\bar{v}}_{i},\vartheta_{q_{3}}^{i})+\frac{1}{2\rho}\sum_{i=1}^{K}\|\mathbf{\bar{v}}_{i}-\mathbf{W}\mathbf{v}_{i}\|^{2}\\
\label{P3b}\text{s.t.} \quad& \eqref{P1b},\eqref{P1c},\eqref{P1d}.
\end{align}
\end{subequations}
Notably, the optimization variables are separated in the constraints, which motivates us to employ the two-layer optimization framework to iteratively solve the problem (P3), where the BCD method is adopted in the inner layer for variable optimization and $\rho$ is updated in the outer layer.

\subsubsection{Inner layer: subproblem with respect to $\{\mathbf{\bar{v}}_{i}\}$} With the fixed $\{\mathbf{W},\mathbf{v}_{i},\vartheta_{q_{3}}^{i}\}$, we can observe that $\mathbf{\bar{v}}_{i}$ and $\mathbf{\bar{v}}_{j}$ ($i\neq j$) are fully independent in the objective function \eqref{P3a}, which indicates that the problem (P3) can be transformed into $K$ independent subproblems without loss of equivalence. The $i$-th subproblem is given by
\begin{subequations}
\begin{align}
\label{P4-1a} (\text{P}4\text{-}1)\  \min\limits_{\mathbf{\bar{v}}_{i},\vartheta_{q_{3}}^{i}}\quad &  \chi_{3}(\mathbf{\bar{v}}_{i},\vartheta_{q_{3}}^{i})+\frac{1}{2\rho}\|\mathbf{\bar{v}}_{i}-\mathbf{W}\mathbf{v}_{i}\|^{2}
\\
\label{P4-1b}  \text{s.t.} \quad &|(\mathbf{a}_{j}^{\varsigma})^{H}\mathbf{\bar{v}}_{i}|^{2} \leq \epsilon,\ i\neq j,
\end{align}
\end{subequations}
To facilitate solving the unconstrained optimization problem (P4-1), we define a new constant matrix, which consists of $Q$ array response vectors, i.e., $\mathbf{A}\triangleq[\mathbf{\bar{a}}(\phi_{1},\varphi_{1},r_{1}),\cdots,\mathbf{a}(\phi_{q_{1}},\varphi_{q_{2}},r_{q_{3}}),\cdots,
\mathbf{\bar{a}}(\phi_{Q_{1}},\varphi_{Q_{2}},r_{Q_{3}})]$. With the arbitrary $\vartheta_{q_{3}}^{i}$, the problem (P4-1) with respect to $\mathbf{\bar{v}}_{i}$ can be formulated as
\begin{subequations}
\begin{align}
\label{P4-2a} (\text{P}4\text{-}2)\  \min\limits_{\mathbf{\bar{v}}_{i}}\quad
&\|\mathbf{t}_{i}-\mathbf{A}\mathbf{\bar{v}}_{i}\|^{2} +\frac{1}{2\rho}\|\mathbf{\bar{v}}_{i}-\mathbf{W}\mathbf{v}_{i}\|^{2}\\
\label{P4-2b}  \text{s.t.} \quad &\eqref{P4-1b},
\end{align}
\end{subequations}
where the $q$-th element of $\mathbf{t}_{i}$ is given by
\begin{align}\label{dec_vec}
\mathbf{t}_{i}^{[q]}=\begin{cases}te^{\jmath\vartheta_{q_{3}}^{i}},\ &\text{if}\ \phi_{q_{1}}=\phi_{i},\ \varphi_{q_{2}}=\varphi_{i},\ r_{q_{3}}\in[r^{\text{N}}_{i},r^{\text{F}}_{i}],\\
 0 &\text{otherwise},
\end{cases}
\end{align}
Problem (P4-2) is a convex programming, where the optimal $\mathbf{\bar{v}}_{i}$ can be directly obtained via the standard convex solver, such as CVX.
\subsubsection{Inner layer: subproblem with respect to $\{\vartheta_{q_{3}}^{i}\}$} Substituting the optimized result of $\mathbf{\bar{v}}_{i}$ into the problem (P4-2), we have
\begin{subequations}
\begin{align}
\label{P4-3a} (\text{P}4\text{-}3)\  &\min\limits_{\vartheta_{q_{3}}^{i}}\quad  \mathbf{t}_{i}^{H}
\mathbf{t}_{i}-2\Re\left\{\mathbf{t}_{i}^{H}\mathbf{A}^{H}\mathbf{\bar{v}}_{i}\right\}\\
 \label{P4-3b}&\text{s.t.} \quad \eqref{dec_vec}.
\end{align}
\end{subequations}
Note that due to the existence of the equality constraint \eqref{dec_vec}, the optimal $\vartheta_{q_{3}}^{i}$ cannot be directly derived through the first-order optimality condition. Instead, we can observe $\mathbf{t}_{i}$ is a sparse vector, which motivates us to reformulate the problem (P4-3) by shortening the $\mathbf{t}_{i}$ without the loss of equivalence. Specifically, let $\mathbf{\bar{t}}_{i}$ denote the subvector of $\mathbf{t}_{i}$ consisting of all the non-zero elements, it can be expressed as
\begin{align}\label{non_zero_t}
\mathbf{\bar{t}}_{i}=\left[\mathbf{t}_{i}^{[q_{i}^{-}]},\mathbf{t}_{i}^{[q_{i}^{-}+1]},\cdots,\mathbf{t}_{i}^{[q_{i}^{+}-1]},\mathbf{t}_{i}^{[q_{i}^{+}]}\right],
\end{align}
where $q_{i}^{-}$ and $q_{i}^{+}$ denote the indexes of the first and last non-zero elements of $\mathbf{t}_{i}$. Thus, we rewrite the objective \eqref{P4-3a} as
\begin{align}\label{obj_rew}
\eqref{P4-3a} = \mathbf{\bar{t}}_{i}^{H}\mathbf{\bar{B}}_{i}\mathbf{\bar{t}}_{i}-2\Re\left\{\mathbf{\bar{t}}_{i}^{H}\mathbf{\bar{c}}_{i}\right\},
\end{align}
where $\mathbf{B}=\mathbf{I}$ and $\mathbf{c}_{i}=\mathbf{A}^{H}\mathbf{\bar{v}}_{i}$. Here, we extract a sub-matrix $\mathbf{\bar{B}}_{i}$ from $\mathbf{B}$, which contains the entries whose column and row index range from $q_{i}^{-}$ to $q_{i}^{+}$. Similarly, $\mathbf{\bar{c}}_{i}$ denotes a sub-vector of $\mathbf{c}_{i}$ with the element index ranging from $q_{i}^{-}$ to $q_{i}^{+}$. To proceed, the problem (P4-3) can be reformulated as the semidefinite relaxation (SDR) form 
\begin{subequations}
\begin{align}
\label{P4-4a} (\text{P}4\text{-}4)\  \min\limits_{\mathbf{\tilde{T}}_{i}}\quad  &\text{Tr}(\mathbf{\bar{T}}_{i}\mathbf{D}_{i})\\
 \label{P4-4b}\text{s.t.} \quad &\mathbf{\tilde{T}}_{i}^{[j,j]}=t^{2}, \ j\in\{1,\cdots,q_{i}^{+}-q_{i}^{-}+1\},\\
  \label{P4-4c} & \text{rank}(\mathbf{\tilde{T}}_{i}) = 1,\\
  \label{P4-4d} & \mathbf{\tilde{T}}_{i}\succeq \mathbf{0},
\end{align}
\end{subequations}
where $\mathbf{D}_{i}$ and $\mathbf{\tilde{T}}_{i}$ are given by
\begin{align}\label{def_D_T}
\mathbf{D}_{i}=\begin{bmatrix}
\mathbf{\bar{B}}_{i}&  -\mathbf{\bar{c}}_{i}\\
-\mathbf{\bar{c}}_{i}^{H}&  0\\
\end{bmatrix},\quad
\mathbf{\tilde{T}}_{i} = \begin{bmatrix}
\,\mathbf{\bar{t}}_{i}\\
1\\
\end{bmatrix}\cdot[\,\mathbf{\bar{t}}_{i}^{H},1].
\end{align}
By removing the rank-one constraint \eqref{P4-4c}, the problem is degenerated into a standard semidefinite program (SDP), which can be optimally solved by the CVX. Then, the rank-one solution $[\,\mathbf{\bar{t}}_{i}^{H},1]$ can be obtained by the Gaussian randomization procedure with at-least $\frac{\pi}{4}$-approximation accuracy \cite{SDR}.

\subsubsection{Inner layer: subproblem with respect to $\{\mathbf{v}_{i}\}$} With the fixed $\{\mathbf{\bar{v}}_{i},\vartheta_{q_{3}}^{i},\mathbf{W}\}$, the problem (P3) can be divided into $K$ subproblems, where the $i$-th subproblem with respect to $\{\mathbf{v}_{i}\}$ is given by
\begin{subequations}
\begin{align}
\label{P5-1a} (\text{P}5\text{-}1)\  \min\limits_{\mathbf{v}_{i}}\quad &  \|\mathbf{\bar{v}}_{i}-\mathbf{W}\mathbf{v}_{i}\|^{2}\\
\label{P5-1b}\text{s.t.} \quad& \eqref{P1b},\eqref{P1d}.
\end{align}
\end{subequations}
To tackle the non-convex constraint \eqref{P1b}, we rewrite the problem (P5-1) as the SDR form
\begin{subequations}
\begin{align}
\label{P5-2a} (\text{P}5\text{-}2)\  \min\limits_{\mathbf{\tilde{V}}_{i}}\quad &  \text{Tr}(\mathbf{\tilde{V}}_{i}\mathbf{E}_{i})\\
\label{P5-2b}\text{s.t.} \quad& \sum_{j=1}^{K}\mathbf{\tilde{V}}_{i}^{[j,j]}=1,\\
\label{P5-2c} & \mathbf{\tilde{V}}_{i} \succeq \mathbf{0},\\
\label{P5-2d} & \text{rank}(\mathbf{\tilde{V}}_{i}) = 1,
\end{align}
\end{subequations}
where $\mathbf{E}_{i}=\begin{bmatrix}\mathbf{W}^{H}\mathbf{W}&  -\mathbf{W}^{H}\mathbf{\bar{v}}_{i}\\ -\mathbf{\bar{v}}_{i}^{H}\mathbf{W}&  0\\ \end{bmatrix}$, $\mathbf{\tilde{V}}_{i} = \begin{bmatrix} \,\mathbf{v}_{i}\\ 1\\
\end{bmatrix}\cdot[\,\mathbf{v}_{i}^{H},1]$, and $\mathbf{V}_{i}$ denotes the submatrix of $\mathbf{\tilde{V}}_{i}$ that consisting of the former $K$-rows and -columns elements. By ignoring the rank-one solution, the optimal $\mathbf{\tilde{V}}_{i}$ with general rank can be directly obtained by solving the SDP problem. Then, we utilize the Gaussian randomization procedure to extract the rank-one solution from the high-rank $\mathbf{\tilde{V}}_{i}$.

\subsubsection{Inner layer: subproblem with respect to $\{\mathbf{W}\}$} With the fixed $\{\mathbf{\bar{v}}_{i},\vartheta_{q_{3}}^{i},\mathbf{v}_{i}\}$, the problem (P3) is converted to
\begin{subequations}
\begin{align}
\label{P6a} (\text{P}6)\  \min\limits_{\mathbf{v}_{i}}\quad &  \sum_{i=1}^{K}\|\mathbf{\bar{v}}_{i}-\mathbf{W}\mathbf{v}_{i}\|^{2}\\
\label{P6b}\text{s.t.} \quad& \eqref{P1c},
\end{align}
\end{subequations}
which is a convex problem and can be directly solved.

\subsubsection{Outer layer: penalty factor update} When the inner layer iteration converges, we update $\rho$ for obtaining the feasible solution of the original problem (P1), which is updated by
\begin{align}\label{update_penalty_factor}
\rho = \frac{\rho}{\tilde{c}},
\end{align}
where $\tilde{c}>1$ denotes the constant update coefficient.

The specific algorithm details are summarized in the \textbf{Algorithm \ref{TLA}}. Let $f(\mathbf{W}^{l},\mathbf{v}_{i}^{l},\mathbf{\bar{v}}_{i}^{l},(\vartheta_{q_{3}}^{i})^{l})$ denote the objective value at the $l$-th inner-layer iteration, it must hold that
\begin{align}
\nonumber &f(\mathbf{W}^{l},\mathbf{v}_{i}^{l},\mathbf{\bar{v}}_{i}^{l},(\vartheta_{q_{3}}^{i})^{l}) \overset{a}{\geq} f(\mathbf{W}^{l},\mathbf{v}_{i}^{l},\mathbf{\bar{v}}_{i}^{l+1},(\vartheta_{q_{3}}^{i})^{l+1})\overset{b}{\geq}   \\ \label{convergence_BCD}
&\!\!\!f(\mathbf{W}^{l},\mathbf{v}_{i}^{l+1},\mathbf{\bar{v}}_{i}^{l+1},(\vartheta_{q_{3}}^{i})^{l+1})\overset{c}{\geq}   f(\mathbf{W}^{l+1},\mathbf{v}_{i}^{l+1},\mathbf{\bar{v}}_{i}^{l+1},(\vartheta_{q_{3}}^{i})^{l+1}),
\end{align}
where the inequality holds as the suboptimal or optimal solutions are guaranteed at steps 4-7. Thus, the proposed two-layer algorithm remains mono-increased over the inner-layer iterations. For the outer layer, when the penalty factor approaches $0$, the equality constraint $\mathbf{\bar{v}}_{i}=\mathbf{W}\mathbf{v}_{i}$ can be satisfied and the feasible solutions of the problem (P1) can be returned.

The complexity of \textbf{\textbf{Algorithm \ref{TLA}}} is generated by the steps 4 to 7. Specifically, in step 4, we update the optimal $\mathbf{\bar{v}}_{i}$ by solving the second-order cone programming (SOCP) program, which causes the complexity of $\mathcal{O}\big((M_{\text{t}})^{3.5}\big)$. In steps 5 and 6, we solve the standard SDP problem to obtain $\vartheta_{q_{3}}^{i}$ and $\mathbf{v}_{i}$, which suffers the complexity of $\mathcal{O}\big((q_{+}-q_{-}+1)^{3.5}\big)$ and $\mathcal{O}\big((K+1)^{3.5}\big)$, respectively. The problem (P6) is a second-order cone programming (SOCP) program, which can be optimally solved with the complexity of $\mathcal{O}\big((KM_{\text{t}})^{3.5}\big)$. Thus, the whole computational complexity to design the beam-steering beamformers for $K$ NOMA groups is given by $\mathcal{O}\big(l_{\text{out}}l_{\text{inner}}(KM_{\text{t}}^{3.5}+K(q_{+}-q_{-}+1)^{3.5}+K(K+1)^{3.5}+(KM_{\text{t}})^{3.5})\big)$, where $l_{\text{out}}$ and $l_{\text{inner}}$ denote the number of outer and inner iterations.

\begin{algorithm}[t]
    \caption{Two-layer algorithm.}
    \label{TLA}
    \begin{algorithmic}[1]
        \STATE{Initialize $\{\mathbf{W},\mathbf{v}_{i}\}$ and set the convergence accuracy $\varepsilon_{1}$ and $\varepsilon_{2}$.}
        \REPEAT
        \REPEAT
        \STATE{ update $\mathbf{\bar{v}}_{i}$ $(1\leq i\leq K)$ by solving the problem (P4-2).}
        \STATE{ update $\vartheta_{q_{3}}^{i}$ $(1\leq i\leq K)$ by solving the problem (P4-4).}
        \STATE{ update $\mathbf{v}_{i}$ $(1\leq i\leq K)$ by solving the problem (P5-2).}
        \STATE{ update $\mathbf{W}$ by solving the problem (P6).}
        \UNTIL{ the objective function converges with the accuracy $\varepsilon_{1}$.}
        \STATE{ update $\rho$ according to \eqref{update_penalty_factor}.}
        \UNTIL{ the penalty term $\sum_{i=1}^{K}\|\mathbf{\bar{v}}_{i}-\mathbf{W}\mathbf{v}_{i}\|^{2}$ falls below $\varepsilon_{2}$.}
    \end{algorithmic}
\end{algorithm}
\subsection{Optimal Power Allocation Strategy}
It readily knows that the optimized large beam-depth beamformer $\mathbf{W}\mathbf{v}_{i}$ radiates almost no power at the location of the $j$-th ($j\neq i$) NOMA group. Thus, the inter-interference between different NOMA groups is efficiently eliminated, which yields the following SINR/signal-to-noise ratio (SNR) expressions, i.e.,
\begin{align} \label{SINR_NU_new}
\gamma_{\text{N}_{i}\rightarrow \text{N}_{i}} = \frac{P_{1,i}g^{\text{N}}_{i}}
{\sigma^{2}}, \quad \gamma_{\text{N}_{i}\rightarrow \text{F}_{i}} = \frac{P_{2,i}g^{\text{N}}_{i}}
{P_{1,i}g^{\text{N}}_{i}+\sigma^{2}},
\end{align}
\begin{align} \label{SINR_FU_new}
\gamma_{\text{F}_{i}\rightarrow \text{F}_{i}} = \frac{P_{2,i}g^{\text{F}}_{i}}
{P_{1,i}g^{\text{F}}_{i}+\sigma^{2}}.
\end{align}
Here, $g^{\text{N}}_{i}=|(\mathbf{h}^{\text{N}}_{i})^{H}\mathbf{W}\mathbf{v}_{i}|^{2}$ and $g^{\text{F}}_{i}=|(\mathbf{h}_{i}^{\text{F}})^{H}\mathbf{W}\mathbf{v}_{i}|^{2}$ denote the channel gains of $\text{NU}_{i}$ and $\text{FU}_{i}$, respectively.

Then, we aim to maximize the spectral efficiency of the network, where an optimization of maximizing the sum achievable rate of the NU and FU is formulated, subject to the constraints of total transmit power budget at the BS, QoS constraint at the users, and SIC decoding constraint. It is given by
\begin{subequations}
\begin{align}
\label{P7a} (\text{P}7)\  \max\limits_{P_{q,i}}\quad   &\sum_{i}\sum_{\varsigma\in\{\text{N},\text{F}\}} R_{\varsigma_{i}\rightarrow \varsigma_{i}}\\
\label{P7b}\text{s.t.} \ & \sum\nolimits_{i=1}^{K}\sum\nolimits_{q=1}^{2}P_{q,i}\leq P_{\text{max}},\\
\label{P7c}  & R_{\varsigma_{i}\rightarrow \varsigma_{i}} \geq R_{\text{QoS}}^{\varsigma,i},\ 1\leq i\leq K,\\
\label{P7d}  & \gamma_{\text{N}_{i}\rightarrow \text{F}_{i}}\geq \gamma_{\text{F}_{i}\rightarrow \text{F}_{i}},\ 1\leq i\leq K,
\end{align}
\end{subequations}
where $R_{\varsigma_{i}\rightarrow \varsigma_{i}}=\log_{2}(1+\gamma_{\varsigma_{i}\rightarrow \varsigma_{i}})$, \eqref{P7b} limits the transmit power lower than the maximal transmit power budget $P_{\text{max}}$; \eqref{P7c} guarantees the achievable rate $R_{\varsigma_{i}\rightarrow \varsigma_{i}}$ is no less than the QoS requirement $R_{\text{QoS}}^{\varsigma,i}$; \eqref{P7d} accounts for the successful SIC decoding constraint; In the following, we consider deriving the optimal power allocation strategy for the problem (P7).

To elaborate, from the expressions of \eqref{SINR_NU_new} and \eqref{SINR_FU_new}, we can observe that the users in each NOMA group can be regarded to perform the single-input single-output (SISO) transmission, with the determined channel gains $g^{\text{N}}_{i}$ and $g^{\text{F}}_{i}$. Thus, the SIC decoding constraint \eqref{P7d} is equivalent to $g^{\text{N}}_{i}\geq g^{\text{F}}_{i}$ \cite{Y.Liu_MIMO_NOMA}. According to the definition of the BPE, the designed beamformers have the relatively same radiated power at the location of the NU and FU in one group, whereas FU suffers a higher large-scale path loss. It indicates $g^{\text{N}}_{i}\geq g^{\text{F}}_{i}$ always holds for the considered network, i.e., the constraint \eqref{P7d} can be neglected in the problem (P7). Then, the problem (P7) can be converted to
\begin{subequations}
\begin{align}
\label{P8-1a} (\text{P}8\text{-}1)\  \max\limits_{P_{q,i}}\quad   &\sum_{i}\sum_{\varsigma\in\{\text{N},\text{F}\}} R_{\varsigma_{i}\rightarrow \varsigma_{i}}\\
\label{P8-1b}\text{s.t.} \ & \eqref{P7b},\eqref{P7c}.
\end{align}
\end{subequations}
To facilitate the optimization of the problem (P8-1), we consider combining the constraint \eqref{P7b} and \eqref{P7c}. Let $P_{i}^{\text{g}}$ denotes the practical power allocated to the $i$-th NOMA group, we can obtain the feasible set of $P_{i}^{\text{g}}$, i.e., $P_{i}^{\text{g}}\in[P_{\text{min},i},P_{\text{max},i}]$, where $P_{\text{max},i}>0$ and $P_{\text{min},i}>0$ denote maximum and minimum power allocated to the $i$-th NOMA group, respectively. From the constraint \eqref{P7c}, it readily knows that the minimum power allocated to the $i$-th NOMA group should at least guarantee the QoS requirement of each user, this means that
\begin{align} \label{QoS_power_NU}
P_{1,i} =  \frac{\gamma_{\text{QoS}}^{\text{N},i}\sigma^{2}}{g^{\text{N}}_{i}}
\end{align}
\begin{align} \label{QoS_power_FU}
P_{2,i} = \frac{\gamma_{\text{QoS}}^{\text{F},i}\sigma^{2}}{g^{\text{F}}_{i}}+
\gamma_{\text{QoS}}^{\text{F},i}P_{1,i},
\end{align}
where $\gamma_{\text{QoS}}^{\varsigma,i}=2^{R_{\text{QoS}}^{\varsigma,i}}-1$. Substituting \eqref{QoS_power_NU} into \eqref{QoS_power_FU}, we can obtain
\begin{align} \label{min_power}
P_{\text{min},i} = \gamma_{\text{QoS}}^{\text{F},i}\sigma^{2}\left(\frac{1}{g^{\text{F}}_{i}}+\frac{\gamma_{\text{QoS}}^{\text{N},i}}{g^{\text{N}}_{i}}\right)+
\frac{\gamma_{\text{QoS}}^{\text{N},i}\sigma^{2}}{g^{\text{N}}_{i}}.
\end{align}
Due to the maximum transmit power constraint of $\sum_{i=1}^{K}P_{i}^{\text{g}}\leq P_{\text{max}}$, it holds that
\begin{align}\nonumber
P_{\text{max},i} = &P_{\text{max}}-\sum_{t=1,t\neq i}^{K}P_{\text{min},t},\\  \label{max_power}
& P_{\text{max}}-\gamma_{\text{QoS}}^{\text{F},i}\sigma^{2}\left(\frac{1}{g^{\text{F}}_{i}}+
\frac{\gamma_{\text{QoS}}^{\text{N},i}}{g^{\text{N}}_{i}}\right)-
\frac{\gamma_{\text{QoS}}^{\text{N},i}\sigma^{2}}{g^{\text{N}}_{i}}.
\end{align}
With the derived expressions above, we can equivalently convert the problem (P8-1) to the following form with the feasible set as the intersection of closed boxes \cite{S.Rezvani_NOMA}
\begin{subequations}
\begin{align}
\label{P8-2a} (\text{P}8\text{-}2)\  \max\limits_{P_{q,i},P_{i}^{\text{g}}}\quad   &\sum_{i}\sum_{\varsigma\in\{\text{N},\text{F}\}} R_{\varsigma_{i}\rightarrow \varsigma_{i}}\\
\label{P8-2b}\text{s.t.} \ & \sum_{q=1}^{2}P_{q,i}=P_{i}^{\text{g}},\ 1\leq i\leq K,\\
\label{P8-2c}& P_{i}^{\text{g}} \in[P_{\text{min},i},P_{\text{max},i}],\ 1\leq i\leq K,\\
\label{P8-2d}& \sum_{i=1}^{K}P_{i}^{\text{g}}\leq P_{\text{max}},
\end{align}
\end{subequations}
For the problem (P8-2), we propose a two-stage power allocation algorithm. The optimal intra-group power allocation $\{P_{q,i}\}$ is derived under the given $\{P_{i}^{\text{g}}\}$ in the first stage, and the optimal inter-group power allocation $\{P_{i}^{\text{g}}\}$ is obtained via the bisection method in the second stage.
\subsubsection{Problem with respect to $\{P_{q,i}\}$} Given any feasible $\{P_{i}^{\text{g}}\}$, the problem (P8-2) is converted to
\begin{subequations}
\begin{align}
\label{P8-3a} (\text{P}8\text{-}3)\  \max\limits_{P_{q,i}}\quad   &\sum_{i}\sum_{\varsigma\in\{\text{N},\text{F}\}} R_{\varsigma_{i}\rightarrow \varsigma_{i}}\\
\label{P8-3b}\text{s.t.} \ & \eqref{P8-2b}.
\end{align}
\end{subequations}
The objective function \eqref{P8-3a} and the power constraint \eqref{P8-2b} for different NOMA groups are fully separated, which motivates us to divide the problem (P8-3) into $K$ subproblems. In each subproblem, we only focus on the intra-group power allocation for two-user NOMA transmission, where the $i$-th subproblem is given by
\begin{subequations}
\begin{align}
\label{P8-4a} (\text{P}8\text{-}4)\  \max\limits_{P_{q,i}}\quad   &\sum_{\varsigma\in\{\text{N},\text{F}\}} R_{\varsigma_{i}\rightarrow \varsigma_{i}}\\
\label{P8-4b}\text{s.t.} \ & \sum_{q=1}^{2}P_{q,i}=P_{i}^{\text{g}}.
\end{align}
\end{subequations}
Note that the problem (P8-4) is a typical sum rate maximization optimization problem for a single-carrier SISO NOMA network, where the optimal power allocation is to allocate the extra power to the best-channel user while maintaining the QoS requirement of the other users. Thus, the optimal power allocation for the $i$-th NOMA group can be determined as follows.
\begin{align} \label{opt_intra_power_F}
P_{2,i} = \frac{\gamma_{\text{QoS}}^{\text{F},i}(\sigma^{2}+P_{i}^{\text{g}}g^{\text{F}}_{i})}
{g^{\text{F}}_{i}+\gamma_{\text{QoS}}^{\text{F},i}g^{\text{F}}_{i}},
\end{align}
\begin{align}\label{opt_intra_power_N}
P_{1,i} =  P_{i}^{\text{g}}-P_{2,i}.
\end{align}

\begin{algorithm}[t]
    \caption{Bisection algorithm for optimal power allocation.}
    \label{Bisection}
    \begin{algorithmic}[1]
        \STATE{Initialize initial $\mu_{\text{lower}}$ and $\mu_{\text{upper}}$. Set a convergence accuracy $\varepsilon_{3}$.}
        \REPEAT
        \STATE{ $\mu=\frac{\mu_{\text{lower}}+\mu_{\text{upper}}}{2}$.}
        \STATE{ update $\tilde{P}_{i}^{\text{g}}$ according to \eqref{opt_inter_power}.}
        \STATE{ \textbf{if} $\sum_{i=1}^{K}\tilde{P}_{i}^{\text{g}}\geq a_{i}P_{\text{max}}-b_{i}$}
        \STATE{  \quad $\mu_{\text{lower}}=\mu$.}
        \STATE{ \textbf{else}}
        \STATE{  \quad $\mu_{\text{upper}}=\mu$.}
        \STATE{ \textbf{end}}
        \UNTIL{ the $|\sum_{i=1}^{K}\tilde{P}_{i}^{\text{g}}-a_{i}P_{\text{max}}+b_{i}|\leq\varepsilon_{3}$.}
    \end{algorithmic}
\end{algorithm}
\subsubsection{Problem with respect to $\{P_{i}^{\text{g}}\}$} Substituting \eqref{opt_intra_power_F} and \eqref{opt_intra_power_N} into the problem (P8-2), then we can obtain
\begin{subequations}
\begin{align}
\label{P8-5a}(\text{P}8\text{-}5)\  \max\limits_{P_{i}^{\text{g}}}\    &\sum_{i=1}^{K}\log_{2}\Bigg(1+\frac{g^{\text{N}}_{i}}{\sigma^{2}}\Bigg(a_{i}P_{i}^{\text{g}}-b_{i}\Bigg)
\Bigg)\\
\label{P8-5b}\text{s.t.} \ & \eqref{P8-2c},\\
\label{P8-5c}&  \sum_{i=1}^{K}P_{i}^{\text{g}}= P_{\text{max}},
\end{align}
\end{subequations}
where $a_{i} = \frac{1}
{1+\gamma_{\text{QoS}}^{\text{F},i}}$ and $b_{i}=\frac{\gamma_{\text{QoS}}^{\text{F},i}\sigma^{2}}
{g^{\text{F}}_{i}+\gamma_{\text{QoS}}^{\text{F},i}g^{\text{F}}_{i}}$. Note that we neglect the constant term $\sum_{i=1}^{K}R_{\text{QoS}}^{\text{F},i}$ in the objective \eqref{P8-5a} as it does not affect the optimization of the solutions. The problem (P8-5) is a standard convex problem with the affine power constraint, which can be optimally solved by the Lagrange dual approach. Let $\tilde{P}_{i}^{\text{g}}=a_{i}P_{i}^{\text{g}}-b_{i}$, the problem (P8-5) can be reformulated as
\begin{subequations}
\begin{align}
\label{P8-6a}(\text{P}8\text{-}6)\  \max\limits_{\tilde{P}_{i}^{\text{g}}}\    &\sum_{i=1}^{K}\log_{2}\Bigg(1+\frac{g^{\text{N}}_{i}\tilde{P}_{i}^{\text{g}}}{\sigma^{2}}\Bigg)\\
\label{P8-6b}\text{s.t.} \ &  \tilde{P}_{i}^{\text{g}} \in[\tilde{P}_{\text{min},i},\tilde{P}_{\text{max},i}],\ 1\leq i\leq K,\\
\label{P8-6c}&  \sum_{i=1}^{K}\tilde{P}_{i}^{\text{g}}= a_{i}P_{\text{max}}-b_{i},
\end{align}
\end{subequations}
where $\tilde{P}_{\text{min},i}=a_{i}P_{\text{min},i}-b_{i}$ and $\tilde{P}_{\text{max},i}=a_{i}P_{\text{max},i}-b_{i}$. The Lagrange dual function of the problem (P8-6) is given by
\begin{align}\nonumber
\mathcal{L}(\tilde{P}_{i}^{\text{g}},\mu) = & \sum_{i=1}^{K}\log_{2}\Bigg(1+\frac{g^{\text{N}}_{i}\tilde{P}_{i}^{\text{g}}}{\sigma^{2}}\Bigg)+
\mu\Big(a_{i}P_{\text{max}}-b_{i}- \\ \label{Lagrange}
&\sum_{i=1}^{K}\tilde{P}_{i}^{\text{g}}\Big),
\end{align}
where $\mu\geq0$ denotes the Lagrange multiplier for the constraint \eqref{P8-6c}. Reviewing the Karush-Kuhn-Tucker (KKT) conditions below
\begin{subequations}
\begin{align}
    \label{KKT-1}\text{K1}: \quad & \frac{\partial \mathcal{L}(\tilde{P}_{i}^{\text{g}},\mu)}{\partial \tilde{P}_{i}^{\text{g}}}=0, \ \forall i, \\
    \label{KKT-2} \text{K2}: \quad &\sum_{i=1}^{K}\tilde{P}_{i}^{\text{g}}=a_{i}P_{\text{max}}-b_{i},\\
    \label{KKT-3} \text{K3}: \quad &\tilde{P}_{i}^{\text{g}} \in[\tilde{P}_{\text{min},i},\tilde{P}_{\text{max},i}],\ 1\leq i\leq K,
\end{align}
\end{subequations}
we can obtain the optimal solutions of $\{\tilde{P}_{i}^{\text{g}}\}$ via the one-dimension search for the Lagrange multiplier $\mu$. In particular, with any given $\mu$, the optimal inter-group power allocation is given by
\begin{align}\label{opt_inter_power}
\tilde{P}_{i}^{\text{g}}=\begin{cases}\left(\frac{g^{\text{N}}_{i}}{\mu\sigma^{2}\ln2}-1\right)\frac{\sigma^{2}}{g^{\text{N}}_{i}},\ \text{if}\  \tilde{P}_{i}^{\text{g}}\in[\tilde{P}_{\text{min},i},\tilde{P}_{\text{max},i}],\\
0,\ \text{otherwise}.
\end{cases}
\end{align}
Here, we use the Bisection algorithm to search for the optimal $\mu$ with the convergence condition of $\sum_{i=1}^{K}\tilde{P}_{i}^{\text{g}}=a_{i}P_{\text{max}}-b_{i}$. The detailed pseudo code is summarized in the \textbf{Algorithm \ref{Bisection}}. As the intra-group power allocation is derived in the closed-form expression, the main computational complexity of optimal power allocation calculation relies on the Bisection algorithm, which suffers a complexity of $\mathcal{O}\Big(\log_{2}(\frac{\mu_{\text{upper}}-
\mu_{\text{lower}}}{\varepsilon_{3}})\Big)$.

\section{beam-splitting beamformer Scheme}\label{beam-splitting}
\begin{figure}[t]
  \centering
  \includegraphics[scale = 0.6]{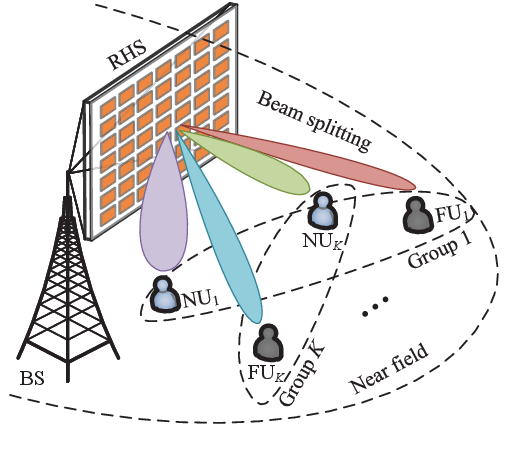}
  \caption{beam-splitting scheme for near-field NOMA transmission.}
  \label{Fig.4}
\end{figure}
In this section, we consider the random user location distribution. As shown in Fig. \ref{Fig.4}, a beam-splitting-based hybrid beamformer strategy is proposed for NOMA transmission, in which each individual user in a single NOMA group is served by a sub-beamformer. Then, the optimal power allocation is derived for the proposed beam-splitting scheme.

\subsection{beam-splitting Beamformer Design}
The idea of beam-splitting is to construct multiple sub-beamformers using one hybrid beamformer, i.e., $\mathbf{W}\mathbf{v}_{i}=\mathbf{W}\mathbf{v}_{i,\text{N}}+\mathbf{W}\mathbf{v}_{i,\text{F}}$, where the sub-beamformers $\mathbf{W}\mathbf{v}_{i,\text{N}}$ and $\mathbf{W}\mathbf{v}_{i,\text{F}}$ are designed to serve the NU and FU of the $i$-th group. The optimization problem can be formulated by
\begin{subequations}
\begin{align}
\label{P9-1a} (\text{P}9\text{-}1)\  \max\limits_{\mathbf{W},\mathbf{v}_{i,\text{N}},\mathbf{v}_{i,\text{F}}}\ &\sum_{i=1}^{K}\left(\min\limits_{\varsigma\in\{\text{N},\text{F}\}} |(\mathbf{a}_{i}^{\varsigma})^{H}\mathbf{W}\mathbf{v}_{i,\varsigma}|^{2}\right) \\
\label{P9-1b}\text{s.t.} \quad & \|\mathbf{v}_{i,\text{N}}+\mathbf{v}_{i,\text{F}}\|^{2} = 1,\ \forall i,\\
\label{P9-1c}  &\mathbf{W}^{[m,n]}\in [0,1], \  \forall m, n,\\
\label{P9-1d}  &|(\mathbf{a}_{i}^{\varsigma})^{H}\mathbf{W}\mathbf{v}_{j,\varsigma}|^{2} \leq \epsilon,\ i\neq j,
\end{align}
\end{subequations}
For fairness guarantee, we focus on a max-min objective in the formulated problem (P9-1), which aims to radiate the same power on the locations of the NU and the FU in the common group. However, the problem (P9-1) is intractable to solve due to the coupled objective function \eqref{P9-1a} and the equation constraint \eqref{P9-1b}. In the following, we propose an AO algorithm to solve it.

\subsubsection{Subproblem with respect to $\{\mathbf{v}_{i,\text{N}},\mathbf{v}_{i,\text{F}}\}$}
With the fixed $\mathbf{W}$, the problem (P9-1) is reduced to
\begin{subequations}
\begin{align}
\label{P9-2a} (\text{P}9\text{-}2)\  \max\limits_{\mathbf{v}_{i,\text{N}},\mathbf{v}_{i,\text{F}}}\ &\sum_{i=1}^{K}\left(\min\limits_{\varsigma\in\{\text{N},\text{F}\}} |(\mathbf{a}_{i}^{\varsigma})^{H}\mathbf{W}\mathbf{v}_{i,\varsigma}|^{2}\right) \\
\label{P9-2b}\text{s.t.} \quad & \eqref{P9-1b},\eqref{P9-1d}.
\end{align}
\end{subequations}
From the problem (P9-2), it readily knows that $\mathbf{v}_{i,\varsigma}$ and $\mathbf{v}_{j,\varsigma}$ ($i\neq j$) are uncoupled, which implies that the problem (P9-2)can be reformulated as $K$ subproblems without loss of equivalence. For the subproblem of the $i$-th NOMA group, the optimal $\mathbf{v}_{j,\varsigma}$ that can maximize the minimum objective function \eqref{P9-2a} is derived by
\begin{align}\label{opt_v_split1}
\mathbf{v}_{i,\varsigma} = \alpha_{i,\varsigma}\mathbf{W}^{H}\mathbf{a}_{i}^{\varsigma},
\end{align}
where the unit-modules constraint is neglected in \eqref{P9-2a}. Note that $\alpha_{i,\text{N}}$ and $\alpha_{i,\text{F}}$ are required to satisfy
\begin{align}\label{opt_v_split2}
\alpha_{i,\text{N}} = \alpha_{i,\text{F}}\frac{\|\mathbf{W}^{H}\mathbf{a}_{i}^{\text{F}}\|^{2}}
{\|\mathbf{W}^{H}\mathbf{a}_{i}^{\text{N}}\|^{2}}.
\end{align}
Recalling the unit-modules constraint, we also have $\|\alpha_{i,\text{N}}\mathbf{W}^{H}\mathbf{a}_{i}^{\text{N}}+\alpha_{i,\text{F}}\mathbf{W}^{H}\mathbf{a}_{i}^{\text{F}}\|
=\left\|\alpha_{i,\text{F}}\frac{\|\mathbf{W}^{H}\mathbf{a}_{i}^{\text{F}}\|^{2}}
{\|\mathbf{W}^{H}\mathbf{a}_{i}^{\text{N}}\|^{2}}
\mathbf{W}^{H}\mathbf{a}_{i}^{\text{N}}+\alpha_{i,\text{F}}\mathbf{W}^{H}\mathbf{a}_{i}^{\text{F}}\right\|=1$
Thus, the optimal $\alpha_{\text{F}}$ is given by
\begin{align}\label{opt_v_split3}
\alpha_{i,\text{F}}= \frac{1}{\left\|\alpha_{i,\text{F}}\frac{\|\mathbf{W}^{H}\mathbf{a}_{i}^{\text{F}}\|^{2}}
{\|\mathbf{W}^{H}\mathbf{a}_{i}^{\text{N}}\|^{2}}
\mathbf{W}^{H}\mathbf{a}_{i}^{\text{N}}+\alpha_{i,\text{F}}
\mathbf{W}^{H}\mathbf{a}_{i}^{\text{F}}\right\|}.
\end{align}
Substituting \eqref{opt_v_split3} into \eqref{opt_v_split2}, we can obtain the optimal $\alpha_{i,\text{N}}$.

\subsubsection{Subproblem with respect to $\{\mathbf{W}\}$} Under the given $\mathbf{v}_{i,\varsigma}$, the problem (P9-1) can be transformed into
\begin{subequations}
\begin{align}
\label{P9-3a} (\text{P}9\text{-}3)\  \max\limits_{\mathbf{W}}\ &\sum_{i=1}^{K}\left(\min\limits_{\varsigma\in\{\text{N},\text{F}\}} \text{Tr}(\mathbf{A}_{i}^{\varsigma}\mathbf{W}\mathbf{V}_{i,\varsigma}\mathbf{W}^{H})\right) \\
\label{P9-3b}\text{s.t.} \quad & \eqref{P9-1c},\eqref{P9-1d},
\end{align}
\end{subequations}
where $\mathbf{A}_{i}^{\varsigma} = \mathbf{a}_{i}^{\varsigma}(\mathbf{a}_{i}^{\varsigma})^{H}$ and $\mathbf{V}_{i,\varsigma}=\mathbf{v}_{i,\varsigma}\mathbf{v}_{i,\varsigma}^{H}$. To tackle the non-convex objective function \eqref{P9-3a}, we consider constructing the linear lower-bound function by using the first-order Taylor expansion, which is given by
\begin{align}\nonumber
\mathfrak{L}_{i,\varsigma}(\mathbf{W}) = & -2\text{Tr}((\mathbf{\bar{W}}^{H}-\mathbf{W}^{H})
\mathbf{A}_{i}^{\varsigma}\mathbf{W}\mathbf{V}_{i,\varsigma})+\\ \label{low_bound_func}
&\text{Tr}(\mathbf{A}_{i}^{\varsigma}\mathbf{\bar{W}}\mathbf{V}_{i,\varsigma}\mathbf{\bar{W}}^{H})\leq
\text{Tr}(\mathbf{A}_{i}^{\varsigma}\mathbf{W}\mathbf{V}_{i,\varsigma}\mathbf{W}^{H}),
\end{align}
where $\mathbf{\bar{W}}$ denotes the value of $\mathbf{W}$ optimized in the previous iteration. Thus, the problem (P9-3) can be efficiently solved by utilizing the SCA technique. The convex subproblem of each SCA iteration is given by
\begin{subequations}
\begin{align}
\label{P9-4a} (\text{P}9\text{-}4)\  \max\limits_{\mathbf{W}}\ &\sum_{i=1}^{K}\left(\min\limits_{\varsigma\in\{\text{N},\text{F}\}} \mathfrak{L}_{i,\varsigma}(\mathbf{W})\right) \\
\label{P9-4b}\text{s.t.} \quad & \eqref{P9-1c},\eqref{P9-1d},
\end{align}
\end{subequations}
which can be directly solved by the CVX toolbox.

\begin{algorithm}[t]
    \caption{AO algorithm for beam-splitting beamformer design.}
    \label{AO}
    \begin{algorithmic}[1]
        \STATE{Initialize initial $\mathbf{\bar{W}}$. Set the convergence accuracy $\varepsilon_{4}$ and $\varepsilon_{5}$.}
        \REPEAT
        \STATE{ update $\mathbf{v}_{i,\text{N}}$ and $\mathbf{v}_{i,\text{F}}$ according to \eqref{opt_v_split1}-\eqref{opt_v_split3}.}
        \REPEAT
        \STATE{ optimize $\mathbf{W}$ by solving the problem (P9-4).}
        \STATE{ update $\mathbf{\bar{W}}=\mathbf{W}$. }
        \UNTIL{ the objective value converges with an accuracy of $\varepsilon_{4}$.}
        \UNTIL{ the objective value converges with an accuracy of $\varepsilon_{5}$.}
    \end{algorithmic}
\end{algorithm}

For the power allocation optimization, it is known that the expressions of the achievable rate of each user are the same as that of the beam-steering case due to the existence of the constraint \eqref{P9-1d}. Thus, the \textbf{Algorithm \ref{Bisection}} can be employed to obtain the optimal power allocation strategy, and we neglected here for brevity.

\section{Numerical Results}\label{Numerical Results}

\begin{figure}[t]
  \centering
  \includegraphics[scale = 0.7]{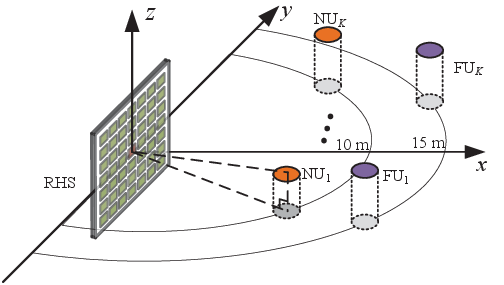}
  \caption{Simulation setup of the considered network.}
  \label{Fig.5}
\end{figure}

\begin{table}[t]
\centering
    \caption{Simulation parameters}\label{tab-1}
    \renewcommand{\arraystretch}{1.4}
\begin{tabular}{|c|c|}
		\hline
        \tabincell{c}{Operating carrier frequency} & $f = 28$ GHz \\
		\hline
        \tabincell{c}{Number of transmit/receive antennas} & $M_{\text{t}} = 1024$\\
		\hline
        \tabincell{c}{Number of equipped  RF chains \\ at the BS} & $M_{\text{t}}^{\text{RF}} = K = 3$\\
		\hline
        \tabincell{c}{Antenna space} & $d = \frac{\lambda}{2}=\frac{c}{2f}$ m \\
		\hline
		\tabincell{c}{Noise power at receivers} & $\sigma^{2}=-75$ dBm \\
		\hline
        \tabincell{c}{QoS requirement} & $R_{\text{QoS}}^{\text{N},i}=R_{\text{QoS}}^{\text{F},j}=R_{\text{QoS}}$\\
		\hline
        \tabincell{c}{Constant scaling coefficients} & $\bar{c}=1.1$\\
        \hline
        \tabincell{c}{Convergence accuracy}
         & \tabincell{c}{$\varepsilon_{1}=\varepsilon_{2}=\varepsilon_{4}=\varepsilon_{5}=10^{-2}$, \\ $\varepsilon_{3}=10^{-6}$} \\
		\hline
        \tabincell{c}{Inter-group interference level}
         & \tabincell{c}{$\epsilon = 10^{-2}$} \\
		\hline
\end{tabular}
\end{table}
The numerical simulation results are provided to validate the effectiveness of the proposed joint beamforming design and power allocation strategies in this section. The simulation setup is depicted in Fig. \ref{Fig.5}. We assume that the central element of the DMA is located at the origin of the coordinate. The NUs and FUs are assumed to be randomly located on the circular rings of 10 meter (m) and 15 m, respectively, where the ranges of the azimuth and elevation angles are from $-\frac{\pi}{2}$ to $\frac{\pi}{2}$. The main simulation parameters are set as Table \ref{tab-1}. Moreover, each figure is the average result of $100$ Monte Carle experiments.

Four baseline schemes are considered in this paper:
\begin{itemize}
\item \textbf{beam-steering/splitting-based FDMA}: In the beam-steering/splitting-based frequency division multiple access (FDMA) scheme, we utilize the proposed beam-steering and beam-splitting schemes to design the spatial beamformers for multiple user groups, where each user group is served by two orthogonal frequency bands of equal size \cite{X.Mu_SIC}. The achievable rate at the NU/FU of the $i$-th group is given by $R_{\varsigma_{i}}=\frac{1}{2}\log_{2}(1+\frac{P_{q,i}|(\mathbf{h}^{\varsigma}_{i})^{H}\mathbf{W}\mathbf{v}_{i}|^{2}}
{\frac{1}{2}\sigma^{2}})$ ($\varsigma_{i}\in\{\text{N}_{i},\text{F}_{i}\}$), where $q=1$ for the NU and $q=2$ for the FU.
  \item \textbf{beam-steering/splitting-based TDMA}: In the beam-steering/splitting-based time division multiple access (TDMA) scheme, the BS serves two users belonging to each group through two equal time slots. Specifically, the BS transmits signals to the NU in the first slot and then the BS communicates with the FU in the second slot. In each time slot, the BS applies the total power of the group to maximize the achievable rate, i.e., $R_{\varsigma_{i}}=\frac{1}{2}\log_{2}(1+\frac{P_{i}^{\text{g}}|(\mathbf{h}^{\varsigma}_{i})^{H}\mathbf{W}\mathbf{v}_{i}|^{2}}
{\sigma^{2}})$.
  \item \textbf{Far-field channel model}: In this scheme, the planar-wave-based far-field channel model is adopted to design the beamformers. Then, we substitute the optimized beamformers into the practical spherical-wave channels to characterize the communication performance of the network.
  \item \textbf{Zero-forcing scheme}: This scheme is a conventional beamforming scheme for the near-field NOMA transmission \cite{NF_NOMA1,NF_NOMA2}, where the zero-forcing (ZF) beamformers are designed only relying on the CSI of NUs, and the FUs are served by the leaked power of the beamformers oriented to the NUs. For simplifying the optimization and without losing conviction, the fully-digital ZF beamformers are considered.
\end{itemize}

\begin{figure}[!t]
 \centering
 \subfigure[Normalized beamforming gain spectrum of the proposed beam-steering scheme.]{
  \includegraphics[scale = 0.46]{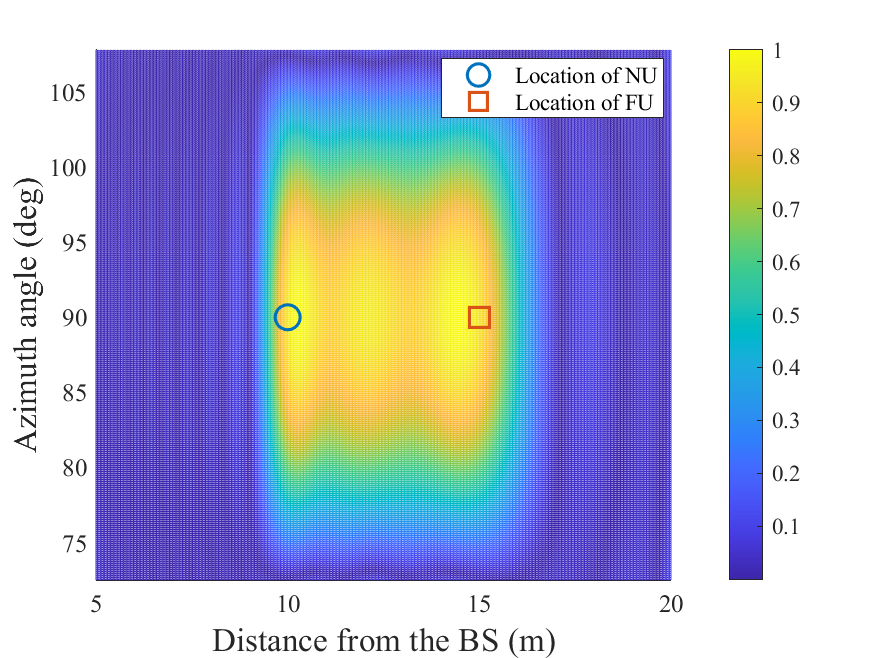}
   \label{Fig.6a}
 }
\subfigure[Normalized beamforming gain spectrum of the proposed beam-splitting scheme.]{
 \includegraphics[scale = 0.46]{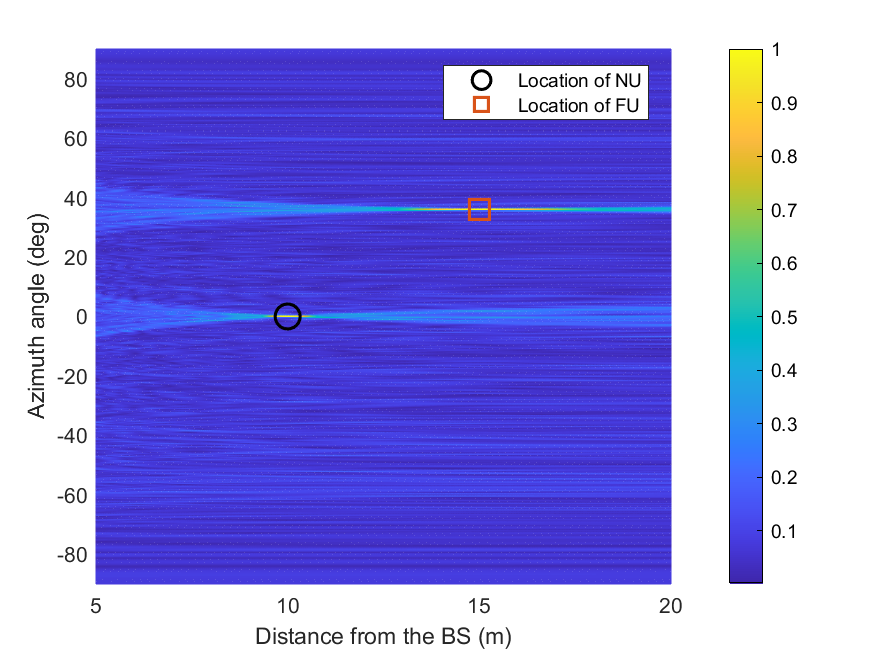}
 \label{Fig.6b}
}
\caption{Radiation pattern of the proposed schemes.}
\label{Fig.6}
\end{figure}

To intuitively illustrate the radiation attributes of the proposed beamforming design schemes in near-field communications, the normalized radiation power spectrums over the free-space location are drawn in Fig. \ref{Fig.6}. We consider the single-group scenario (i.e., $K=1$) with location topology that both the NU and the FU NU and FU lie in the common spatial plane with the same elevation angle $\phi_{\text{N}} = \phi_{\text{F}} = 0^{\circ}$. It can be observed from Fig. \ref{Fig.6a} that the proposed beam-steering scheme can achieve signal power strengthening in the area between the NU and the FU while suppressing signal leakage in other regions of no interest. Meanwhile, Fig. \ref{Fig.6b} shows that the proposed beam-splitting scheme is able to achieve the signal power focus on the multiple locations (also referred as to multi-focus) of the NU and the FU. Both results demonstrate the effectiveness of the proposed beamforming design schemes.

\begin{figure}[t]
  \centering
  \includegraphics[scale = 0.46]{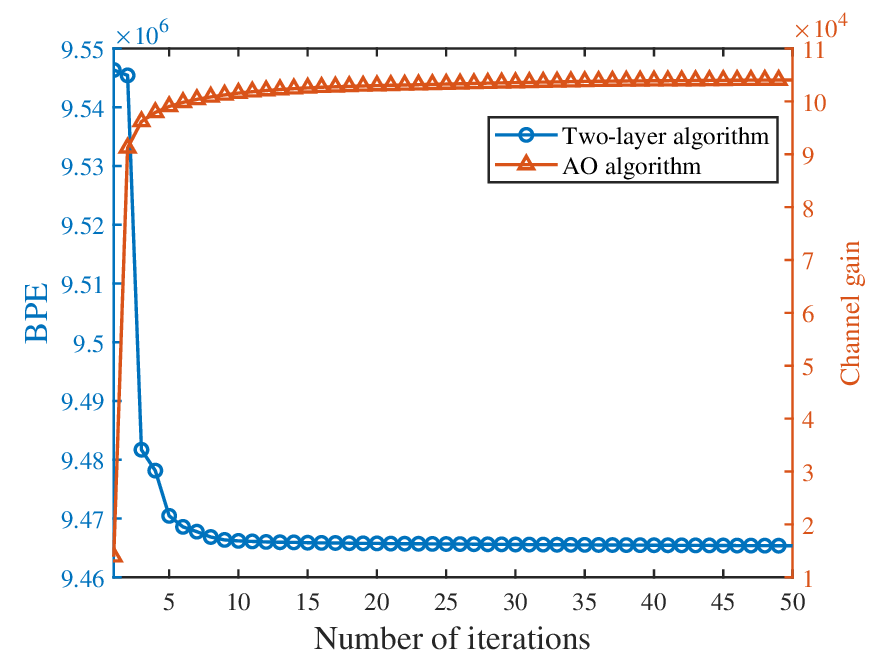}
  \caption{Convergence performance of the proposed algorithms with $t=10^3$, $r_{\text{N},i}=10$ m, and $r_{\text{F},i}=15$ m.}
  \label{Fig.7}
\end{figure}

Fig. \ref{Fig.7} shows the convergence performance of the two-layer and AO algorithms versus the number of iterations. It can be observed that the BPE value monotonically decreases over the iterations and converges to a stable solution in around 10 steps. It is worth noting that the value of BPE does not fall below a very small value at convergence, such as a value close to 0. However, this is to be expected because: 1) BPE incorporates all the error accumulation between the designed beam and the ideal beam characterized by the 3D codebook; and 2) the practical beam needs to approach the ideal beam with a main lobe magnitude of $t^2=10^{6}$, which inevitably leads to a high radiated power leakage in other regions, e.g., the beam pattern also produces a high radiated power in the vicinity of a $90^{\circ}$ azimuth angle in Fig. \ref{Fig.6a}. Also can be seen, the proposed AO algorithm can converge within the finite iterations, which guarantees a high channel gain even at the far users.

\begin{figure}[t]
  \centering
  \includegraphics[scale = 0.46]{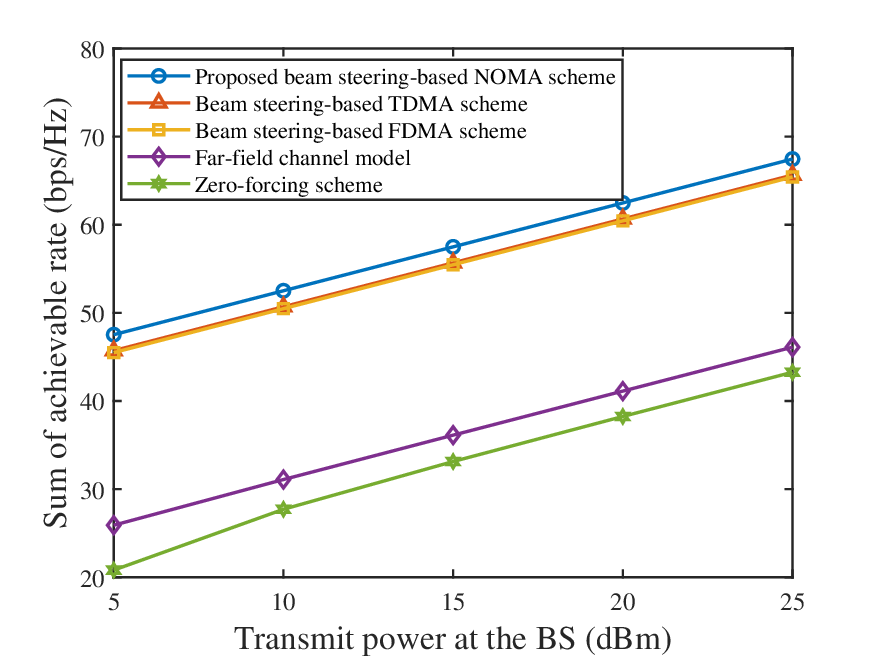}
  \caption{Sum achievable rate versus the transmit power at the BS with $t=10^6$, $r_{\text{N},i}=10$ m, $r_{\text{F},i}=15$ m, and $R_{\text{QoS}}=1$ bps/Hz.}
  \label{Fig.8}
\end{figure}

\begin{figure}[t]
  \centering
  \includegraphics[scale = 0.46]{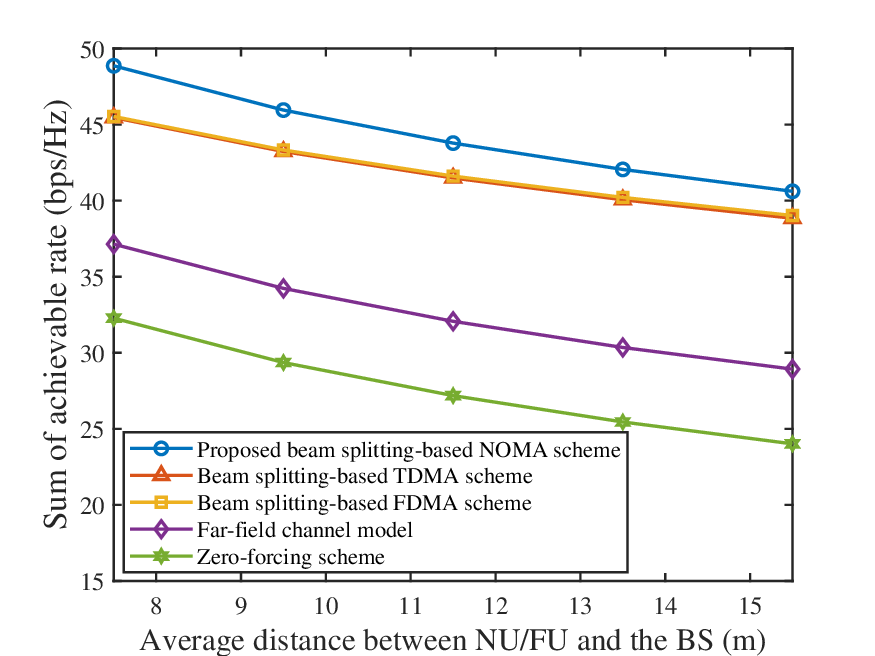}
  \caption{Sum achievable rate versus the average distance from the NU/FU to the BS with $R_{\text{QoS}}=1$ bps/Hz and $P_{\text{max}} = 15$ dBm.}
  \label{Fig.9}
\end{figure}
In Fig. \ref{Fig.8} and Fig. \ref{Fig.9}, we compare the communication performance of the proposed scheme with the other four baseline schemes. Thereinto, the variations of the transmit power and the average distance from the NU and FU to the BS are considered, respectively, where we assume that the NU and FU are apart by a fixed distance of 5 m, and we consider simultaneously changing their positions to adjust average distance of the NU and FU from the BS. It is observed that the proposed beam-steering-based NOMA scheme realizes the highest sum achievable rate among all the schemes. This can be explained by the fact that: 1) NOMA allows the BS to serve the NU and FU in the same time-frequency resource block by flexible power control, which is capacity-achieving and consequently enables better performance than the OMA schemes (i.e., FDMA and TDMA); 2) due to the mismatch between the beam pattern based on the far-field planar-wave channel model and the practical near-field spherical-wave channels, the signal power received at the users are significantly degraded, which deteriorates communication performance of the network; and 3) since the conventional ZF scheme is designed based only on the CSI of the NU, it inevitably leads to an extremely weak channel for FU. Thus, more power should be allocated to the FU to satisfy its QoS requirement, while NU will obtain less power, which limits the communication rate of the network. Moreover, it can be found that the sum achievable rate shows a downward trend with the increasing transmission distance, which is because that a larger transmission distance brings a larger path loss, which acquires more transmit power to maintain the same rate, otherwise the rate will decrease.

\begin{figure}[t]
  \centering
  \includegraphics[scale = 0.46]{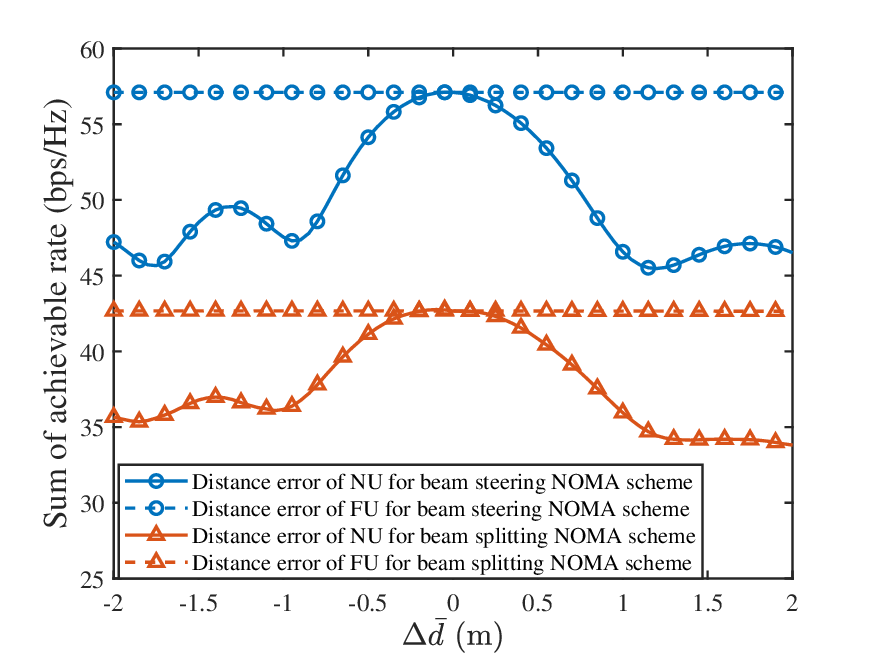}
  \caption{Sum achievable rate versus the distance estimation error of the CSI of the NU and FU with $R_{\text{QoS}}=1$ bps/Hz, $t=10^3$, and $P_{\text{max}} = 15$ dBm.}
  \label{Fig.10}
\end{figure}

Fig. \ref{Fig.10} illustrates the impact of the estimation error of the distance knowledge of the NU and FU on the communication performance of the NOMA network. We assume that all the NUs and FUs are estimated to be located in the distances of 10 m and 15 m, where the realistic distances of NUs and FUs vary from 8 m to 12 m and from 13 m to 17 m, respectively. The distance estimation error for the NU/FU of the $i$-th group is defined by $\Delta d_{\varsigma,i}=d_{\varsigma,i}^{\text{est}}-d_{\varsigma,i}^{\text{rea}}$, $\varsigma\in\{\text{N},\text{F}\}$, where $d_{\varsigma,i}^{\text{est}}$ is the estimated distance and$ d_{\varsigma,i}^{\text{rea}}$ denotes the realistic distance. Then, the average distance estimation error is given by $\Delta \bar{d} = \frac{\sum_{i=1}^{K}\sum_{\varsigma\in\{\text{N},\text{F}\}}\Delta d_{\varsigma,i}}{2K}$. It can be observed that the sum achievable rate of the NOMA network decreases with an increase in $|\Delta \bar{d}|$, which can be expected as the imperfect distance knowledge will cause the mismatch between the beam pattern and the practical spherical-wave channels, thus leading to reduced network performance. Also, we can find an interesting result that both the proposed beamforming design schemes are sensitive to the distance estimation error of the NU while not affected by the imperfect distance information of the FU. This is because, under the optimal power allocation policy, the FU only needs the power to satisfy its own QoS requirement, while all remaining power is allocated to the NU to maximize the rate of that NOMA group, i.e., the achievable rate of the NU dominates the total rate of its NOMA group. Therefore, the degradation of the channel gains of NUs due to imperfect distance information can significantly affect the achievable rate of the network.

\section{Conclusion}\label{Conclusion}
A DMA-enabled near-field NOMA transmission framework was proposed, where NOMA is exploited to enhance the transmission connectivity of the overloaded network. A beam-steering beamforming scheme was proposed for the case of same-direction user distribution, where the BPE metric was introduced to characterize the gap between the hybrid beamformers and desired perfect beamformers. A two-layer algorithm was proposed to minimize the BPE by jointly optimizing the amplitude coefficients of DMA elements and baseband digital beamformers. On this basis, the globally optimal power allocation strategy was obtained according to the KKT conditions. Then, a beam-splitting scheme was proposed for the case of randomly distributed users. An AO algorithm was proposed to generate the sub-beamformers to serve multiple users, where the optimal power allocation is derived for the preconfigured sub-beamformers. It was unveiled that: 1) the proposed beam design schemes show better communication performance than other baseline schemes; 2) the proposed DMA-enabled near-field NOMA transmission framework is sensitive to the distance estimation error of CSI of NUs while insensitive to that of FUs.

\end{document}